\documentclass[journal]{IEEEtran}
\IEEEoverridecommandlockouts

\usepackage{amsmath,amssymb,amsfonts,bm,mathtools,amsthm}
\usepackage{graphicx}
\usepackage[caption=false,font=footnotesize]{subfig}
\usepackage{booktabs,multirow,array}
\usepackage{algorithm}
\usepackage{algorithmic}
\usepackage{xcolor}
\usepackage{cite}
\usepackage{url}

\usepackage[normalem]{ulem}

\usepackage{tikz}
\usetikzlibrary{positioning,calc,arrows.meta,patterns,shapes.misc,decorations.pathreplacing}

\newtheorem{theorem}{Theorem}

\newcommand{\DAZSMR}{\textcolor{black}{\mbox{DAZS-MR}}}
\newcommand{\DAZSPMR}{\textcolor{black}{\mbox{DAZS-PMR}}}
\newcommand{\DAZSMRmath}{\textcolor{black}{\tiny \mbox{DAZS-MR}}}
\newcommand{\DAZSPMRmath}{\textcolor{black}{\tiny \mbox{DAZS-PMR}}}
\newcommand{\cLSFD}{\mbox{c-LSFD}}
\newcommand{\dLSFD}{\mbox{d-LSFD}}
\newcommand{\hLSFD}{\mbox{h-LSFD}}
\newcommand{\pLSFD}{\mbox{p-LSFD}}

\newcommand{\diag}{\operatorname{diag}}
\newcommand{\SINR}{\operatorname{SINR}}
\newcommand{\Zeta}{\mathrm{Z}}
\newcommand{\SE}{\operatorname{SE}}

\begin{document}
\title{Decentralized Decoding with Adaptive Local Combining for Cell-Free Massive MIMO}
\author{\IEEEauthorblockN{Mohd Saif Ali Khan*, Karthik R.M.$^+$ and Samar Agnihotri*}
\thanks{M.~S.~A.~Khan and S.~Agnihotri are with the Indian Institute of Technology Mandi, HP, India (e-mail: d21013@students.iitmandi.ac.in, samar@iitmandi.ac.in). Karthik~R.~M. is with Ericsson India Private Limited, Chennai, TN, India (e-mail: karthik.r.m@ericsson.com).}
\thanks{ The part of this work is published in the International Conference on Signal Processing and Communications (SPCOM), 2026~\cite{khan2026spcom}.}
}

\maketitle

\begin{abstract}
In existing literature, a two-stage architecture comprising of large-scale fading decoding (LSFD) at the central processing unit (CPU) and local combining at the access points (APs) is commonly deployed for uplink processing in scalable cell-free massive multiple-input multiple-output (CF-mMIMO) systems. However, centralized LSFD (\cLSFD) requires cross-AP channel statistics at the CPU and local combining uses schemes that either do not suppress interference optimally or result in analytically intractable solutions. We provide novel schemes to address both concerns. First, we propose the decentralized LSFD (\dLSFD) scheme that maximizes a surrogate signal-to-interference-plus-noise ratio built only from local interference covariance statistics and identify covariance-action regime where it is equivalent to \cLSFD. Second, based on the fact that the antennas at each AP provide spatial degrees of freedom (DoFs) for both interference suppression and coherent array gain, we interpret every combining scheme as solving the optimal DoF allocation problem over available DoFs. Selective zero-forcing (ZF) schemes solve this problem over both the dimension and composition of the local ZF subspace. We then provide the DoF-adaptive ZF-subspace (DAZS) combining scheme, which lets each AP select the optimal directions in the ZF subspace by maximizing its local sum spectral efficiency. Our numerical results establish the efficacy and efficiency of the proposed schemes compared to the corresponding existing schemes.
\end{abstract}


\section{Introduction}
\IEEEPARstart{C}{ell-free} massive multiple-input multiple-output (CF-mMIMO) networks employ geographically distributed access points (APs) to serve user equipment (UE) over shared resources~\cite{nayebi2016performance,ngo2017cell,khan2024joint}, providing macro-diversity and more uniform service than co-located arrays.  Centralized and partially centralized uplink processing in CF-mMIMO can provide strong interference suppression, but their computational costs grow with the amount of signal and statistical information collected across APs~\cite{maryopi2019uplink,bjornson2019making,du2021cell,wang2021partial,khan2024distributed}. To reduce these costs, a scalable uplink processing architecture is proposed~\cite{nayebi2016performance} to perform uplink detection using data decoding at the central processing unit (CPU) using local data estimates formed at the APs~\cite{demir2021foundations}. This architecture presents two distinct design problems: how to construct the decoding weights with limited channel coordination (the decoding problem) and how each AP should form its local estimate to balance interference suppression against coherent array gain (the combining problem).

For the decoding problem, centralized large-scale fading decoding (\cLSFD) is signal-to-interference-plus-noise ratio (SINR)-optimal~\cite{demir2021foundations}. However, computing it requires cross-AP channel statistics and a covariance matrix inversion for each UE at the CPU~\cite{demir2021foundations}. Partial LSFD reduces the set of modeled interferers but retains centralized weight construction~\cite{demir2021foundations,chen2020structured}. Fully decentralized alternatives avoid this coordination by designing the local scalar weights using large-scale fading coefficients (LSFCs)~\cite{schulz2024scalable}. However, two issues remain unresolved: Which decentralized decoding rule is optimal under an explicit local-information restriction, and which part of the discarded cross-AP information accounts for performance loss relative to \cLSFD? It is essential to address these issues as fronthaul signaling and processing can constitute a significant component of the network energy costs~\cite{khan2025energy}.

The combining problem has its own challenges. At each AP, the number of antennas provide spatial degrees of freedom (DoFs), which serve both coherent array gain and interference suppression: every interference direction that is explicitly suppressed consumes one spatial DoF that may otherwise contribute to array gain. Maximum ratio (MR) combining uses no DoF for suppression, whereas minimum mean square error (MMSE)-based combining suppresses interference softly over all DoFs~\cite{ngo2017cell,bjornson2020scalable,polegre2021pilot}; but, neither makes an explicit allocation of spatial DoFs. Local zero-forcing (ZF) schemes offer an intermediate design point by assigning some of the DoFs to explicit suppression. For such schemes the central design question is: \emph{how many} and \emph{which} directions to suppress while preserving sufficient DoFs for array gain? Partial full-pilot ZF (PFZF) and protective weak PFZF (PWPFZF) answer this with a predetermined LSFC-based grouping rule~\cite{zhang2021local}. This rule ranks UE by large-scale gain alone, but does not weigh the interference removed against the array gain lost. Consequently, neither the dimension nor the composition of the ZF subspace is treated as a design variable, generally leading to suboptimal allocation and resulting performance loss. Fig.~\ref{fig:dazs_concept} illustrates these DoF allocation schemes at a single AP.

\begin{figure*}[!t]
\centering
\definecolor{dzDes}{RGB}{31,119,180}   
\definecolor{dzSup}{RGB}{214,39,40}    
\definecolor{dzGain}{RGB}{255,187,120} 
\definecolor{dzSoft}{RGB}{148,103,189} 
\definecolor{dzComb}{RGB}{44,160,44}   
\def\dazsX#1{%
  \draw[xmk] ($#1+(-0.10,-0.10)$) -- ($#1+(0.10,0.10)$);
  \draw[xmk] ($#1+(-0.10,0.10)$)  -- ($#1+(0.10,-0.10)$);}
\def\dazsGeom#1#2#3#4{%
  \coordinate (AP) at (2.1,3.40);
  \foreach \a/\n/\r/\s/\an/\dx/\dy in {175/1/1.55/#1/south/0.05/0.05,
                                      115/2/1.45/#2/south east/0.02/0.00,
                                       40/3/1.30/#3/south west/0.02/0.00,
                                       15/4/0.90/#4/north west/-0.37/-0.16}{%
    \draw[\s] (AP) -- ++(\a:\r);
    \node[clbl, anchor=\an] at ($(AP)+(\a:\r)+(\dx,\dy)$) {$\hat{\mathbf{g}}_{mk_{\n}}$};}
  \draw[tgt] (AP) -- ++(90:1.60);
  \node[clbl, anchor=south, text=dzDes!85!black] at ($(AP)+(90:1.62)$) {$\hat{\mathbf{g}}_{mt}$};
  \fill (AP) circle (1.6pt);
  \node[clbl, anchor=north] at ($(AP)+(0,-0.07)$) {AP $m$};}
\def\dazsV#1{%
  \draw[comb] (AP) -- ++(#1:1.40);
  \node[clbl, text=dzComb!75!black, anchor=south west] at ($(AP)+(#1:1.40)+(0.03,0)$) {$\mathbf v_{mt}$};}
\def\dazsRel#1{\node[font=\scriptsize] at (2.1,2.82) {#1};}
\def\dazsBar#1{%
  \foreach \t [count=\i from 0] in {#1}
    {\draw[\t] (0.85+\i*0.50,2.27) rectangle ++(0.50,0.30);}}

\def\dazsAxes#1#2{%
  \begin{scope}[shift={(0.80,0.85)}]
    \draw[-{Latex[length=1.2mm]}, black!70, line width=0.5pt] (0,0) -- (2.75,0)
      node[right,font=\scriptsize,inner sep=1pt] {DoF};
    \draw[-{Latex[length=1.2mm]}, black!70, line width=0.5pt] (0,0) -- (0,0.86)
      node[left,font=\scriptsize,inner sep=2pt] {SE};
    \foreach \L in {0,...,5}{%
      \draw[black!70] ({0.15+0.46*\L},0) -- ++(0,-0.05);
      \ifnum\L=#1\relax
        \node[font=\scriptsize\bfseries,text=#2,inner sep=1pt] at ({0.15+0.46*\L},-0.17) {\L};
      \else
        \node[font=\scriptsize,inner sep=1pt] at ({0.15+0.46*\L},-0.17) {\L};
      \fi}
  \end{scope}}

\def\dazsCurve{%
  \begin{scope}[shift={(0.80,0.85)}]
    \draw[black!80, line width=0.9pt] plot[smooth, tension=0.6] coordinates
      {(0.15,0.38) (0.61,0.52) (1.07,0.68) (1.53,0.60) (1.99,0.40) (2.45,0.16)};
  \end{scope}}

\def\dazsPick#1#2#3{%
  \begin{scope}[shift={(0.80,0.85)}]
    \draw[#3, densely dotted, line width=0.7pt] (#1,#2) -- (#1,0);
    \fill[#3] (#1,0) circle (1.9pt);
    \fill[#3] (#1,#2) circle (2.6pt);
  \end{scope}}
\resizebox{\textwidth}{!}{%
\begin{tikzpicture}[
  font=\footnotesize,
  panel/.style={draw=black!65, rounded corners=3pt, line width=0.6pt},
  prop/.style ={draw=dzDes!80!black, fill=dzDes!6, rounded corners=3pt, line width=1.2pt},
  kept/.style ={-{Latex[length=2mm]}, line width=0.9pt, black!45},
  supp/.style ={-{Latex[length=2mm]}, line width=1.2pt, dzSup},
  soft/.style ={-{Latex[length=2mm]}, line width=0.8pt, densely dashed, black!65},
  tgt/.style  ={-{Latex[length=2.6mm]}, line width=1.7pt, dzDes!90!black},
  comb/.style ={-{Latex[length=2.4mm]}, line width=1.4pt, dzComb!85!black},
  lbl/.style  ={font=\footnotesize, inner sep=1pt},
  clbl/.style ={font=\scriptsize, inner sep=1pt},
  xmk/.style  ={line width=1.4pt, dzSup!75!black},
  cd/.style   ={draw=black!80, line width=0.5pt, fill=dzDes},
  cs/.style   ={draw=black!80, line width=0.5pt, fill=dzSup!15,
                postaction={pattern=north east lines, pattern color=dzSup}},
  cg/.style   ={draw=black!80, line width=0.5pt, fill=dzGain},
  ttl/.style  ={font=\small\bfseries},
  rul/.style  ={font=\scriptsize\itshape, align=center, text width=3.9cm},
  ptl/.style  ={font=\scriptsize, inner sep=1pt},
  sep/.style  ={black!25, line width=0.5pt}
]

\begin{scope}[shift={(0,0)}]
  \draw[panel] (0,0) rectangle (4.2,5.85);
  \node[ttl] at (2.1,5.58) {(a) MR};
  \dazsGeom{kept}{kept}{kept}{kept}
  \draw[comb] (AP) -- ++(90:1.30);
  \node[clbl, text=dzComb!75!black, anchor=west] at ($(AP)+(0.08,1.10)$) {$\mathbf v_{mt}$};
  \dazsRel{$\mathbf v_{mt}\parallel\hat{\mathbf g}_{mt}$: no suppression}
  \dazsBar{cg,cg,cg,cg,cg}
  \node[font=\fontsize{6.3}{7}\selectfont] at (2.1,2.09) {0 DoF ZF Subspace};
  \draw[sep] (0.35,1.85) -- (3.85,1.85);
  \dazsAxes{0}{black}
  \dazsCurve
  \dazsPick{0.15}{0.38}{black!85}
  \node[ptl, anchor=north west] at (0.80+0.24,0.85+0.30) {no suppression};
  \node[rul] at (2.1,0.28) {Array gain only and no DoF allocation };
\end{scope}

\begin{scope}[shift={(4.5,0)}]
  \draw[panel] (0,0) rectangle (4.2,5.85);
  \node[ttl] at (2.1,5.58) {(b) LP-MMSE};
  \fill[dzSoft, opacity=0.18]
    (2.1,3.40) -- ++(15:1.62) arc[start angle=15, end angle=175, radius=1.62] -- cycle;
  \dazsGeom{soft}{soft}{soft}{soft}
  \dazsV{70}
  \dazsRel{$\mathbf v_{mt}$: soft suppression}
  \shadedraw[left color=dzSoft!28, right color=dzSoft!28, draw=black!80,
             line width=0.5pt] (0.85,2.27) rectangle ++(2.50,0.30);
  \node[font=\fontsize{6.3}{7}\selectfont] at (2.1,2.09) {all 5 DoF, soft split};
  \draw[sep] (0.35,1.85) -- (3.85,1.85);
  \dazsAxes{5}{dzSoft!80!black}
  \dazsPick{2.45}{0.74}{dzSoft!85!black}
  \node[ptl, anchor=east, text=dzSoft!80!black] at (0.80+2.36,0.85+0.74) {all DoF};
  \node[rul] at (2.1,0.28) {Soft weighting over all DoF};
\end{scope}

\begin{scope}[shift={(9.0,0)}]
  \draw[panel] (0,0) rectangle (4.2,5.85);
  \node[ttl] at (2.1,5.58) {(c) PFZF / PWPFZF};
  \draw[black!60, line width=0.8pt, dash pattern=on 3pt off 2pt]
    ($(2.1,3.40)+(180:1.20)$) arc[start angle=180, end angle=0, radius=1.20];
  \node[font=\scriptsize\itshape, black!75, anchor=north] at (0.90,3.34) {LSFC threshold};
  \dazsGeom{supp}{supp}{supp}{kept}
  \dazsV{58}
  \dazsX{($(AP)+(175:0.85)$)}
  \dazsX{($(AP)+(115:0.85)$)}
  \dazsX{($(AP)+(40:0.85)$)}
  \dazsRel{$\mathbf v_{mt}\perp\hat{\mathbf g}_{mk_{1,2,3}}$;\; $\hat{\mathbf g}_{mt}$ preserved}
  \dazsBar{cd,cs,cs,cs,cg}
  \node[font=\fontsize{6.3}{7}\selectfont] at (2.1,2.09) {4 DoF ZF Subspace \,$\cdot$\, 1 DoF Gain};
  \draw[sep] (0.35,1.85) -- (3.85,1.85);
  \dazsAxes{4}{dzSup}
  \dazsCurve
  \draw[black!55, line width=0.6pt, fill=white] (0.80+1.07,0.85+0.68) circle (2.2pt);
  \dazsPick{1.99}{0.40}{dzSup}
  \node[ptl, anchor=west, align=left, text=dzSup!85!black] at (0.80+2.08,0.85+0.62) {LSFC\\threshold};
  \node[rul] at (2.1,0.28) {Allocate DoF based on LSFC threshold};
\end{scope}

\begin{scope}[shift={(13.5,0)}]
  \draw[prop] (0,0) rectangle (4.2,5.85);
  \node[ttl] at (2.1,5.58) {(d) DAZS (proposed)};
  \dazsGeom{kept}{supp}{kept}{kept}
  \dazsV{74}
  \dazsX{($(AP)+(115:0.85)$)}
  \dazsRel{$\mathbf v_{mt}\perp\hat{\mathbf g}_{mk_{2}}$;\; $\hat{\mathbf g}_{mt}$ preserved}
  \dazsBar{cd,cs,cg,cg,cg}
  \node[font=\fontsize{6.3}{7}\selectfont] at (2.1,2.09) {2 DoF ZF Subspace \,$\cdot$\, 3 DoF Gain};
  \draw[sep] (0.35,1.85) -- (3.85,1.85);
  \dazsAxes{2}{dzDes}
  \dazsCurve
  \dazsPick{1.07}{0.68}{dzDes}
  \node[ptl, anchor=south, text=dzDes!85!black] at (0.80+1.42,0.85+0.74) {max local sum SE};
  \node[rul] at (2.1,0.28) {Allocate DoF based on optimization};
\end{scope}
\node[anchor=north, inner sep=0pt] at (8.85,-0.22) {%
  \tikz[baseline=-0.6ex]\draw[tgt] (0,0) -- (0.55,0);~desired $\hat{\mathbf g}_{mt}$\quad
  \tikz[baseline=-0.6ex]\draw[comb] (0,0) -- (0.55,0);~combiner $\mathbf v_{mt}$\quad
  \tikz[baseline=-0.6ex]\draw[cs] (0,-0.12) rectangle (0.34,0.12);~ZF subspace DoF\quad
  \tikz[baseline=-0.6ex]\draw[cg] (0,-0.12) rectangle (0.34,0.12);~array gain DoF\quad
  \tikz[baseline=-0.6ex]{\draw[supp] (0,0) -- (0.55,0);
    \draw[xmk] (0.175,-0.075) -- (0.325,0.075); \draw[xmk] (0.175,0.075) -- (0.325,-0.075);}~ZF suppression\quad
  \tikz[baseline=-0.6ex]\fill[dzSoft, opacity=0.22] (0,-0.13) -- ++(0:0.36) arc[start angle=0, end angle=90, radius=0.36] -- cycle;~soft suppression};
\end{tikzpicture}}
\caption{Spatial-DoF allocation at AP $m$ ($\text{DoF}=5$). Arrow length indicates LSFC; the combining vector is $\mathbf v_{mt}$, and the desired direction is $\hat{\mathbf{g}}_{mt}$. SE curves are schematic.}
\label{fig:dazs_concept}
\end{figure*}

This paper addresses both problems under a single principle: each AP makes its decoding and combining decisions using only the channel statistics available to it. First,
 we derive decentralized LSFD (\dLSFD) as a rule that maximizes a SINR-surrogate constructed from the interference information available at each AP. Using the resulting weights
 the CPU only combines weighted soft estimates, without cross-AP covariance exchange or per-UE matrix inversion. 
We then identify the covariance-action regimes in which the discarded covariance has little impact on the SINR relative to \cLSFD\ and consequently \dLSFD\ is equivalent to \cLSFD. Second,
 we provide a novel interpretation of selective ZF as the optimal spatial DoF allocation problem over both the dimension and composition of the local ZF subspace. The proposed DoF-adaptive ZF-subspace (DAZS) combining
jointly optimizes both the dimension and composition of the local ZF subspace to maximize the local sum spectral efficiency (SE), instead of fixing those by a predetermined threshold, as in \cite{zhang2021local}. It therefore suppresses a direction only when the suppression outweighs the array-gain loss and reduces to MR when suppression does not pay off.

\subsection{Related Work}
LSFD is a second-stage receiver for uplink CF-mMIMO, where locally combined signals are fused using second-order channel statistics~\cite{demir2021foundations}. Scalable variants reduce the number of APs or modeled interferers involved in this processing~\cite{chen2020structured}. In~\cite{zheng2022team,zhang2026on}, the authors reduce different components of centralized processing, but generally retain some form of network-level information exchange.  In~\cite{schulz2024scalable}, the authors present LSFC-based local LSFD weights without any theoretical guarantees.  In~\cite{ma2026fully}, the proposed architecture removes the CPU and obtains coordinated solutions through iterative information exchange among APs.  In contrast, the present work retains CPU-based data fusion but decentralizes the construction of the LSFD weights under an explicit local-information constraint, requiring neither cross-AP covariance exchange nor iterative inter-AP consensus. We further characterize the relation between \cLSFD\ and the proposed \dLSFD.

Local combining in CF-mMIMO spans a wide range of interference-suppression capabilities. MR combining preserves array gain but does not suppress inter-user interference~\cite{ngo2017cell}. LP-MMSE combining provides stronger interference suppression~\cite{bjornson2020scalable}, however, it does not provide a tractable closed-form SE expression, making system-level analysis and optimization challenging. Related receiver designs also exploit distributed channel statistics to mitigate coherent pilot-contaminating interference through reduced-complexity MMSE and generalized-MR processing~\cite{polegre2021pilot}. ZF-based receivers provide an intermediate design by explicitly using spatial DoFs to null selected interference directions. Full ZF and full-pilot ZF suppress all feasible directions, whereas partial ZF variants reduce the nulling dimension to preserve array gain and scalability~\cite{zhang2021local}. In particular, PFZF and PWPFZF construct reduced ZF subspaces from LSFCs-based UE grouping rules~\cite{zhang2021local,khan2025comments}.  In these works, the composition of the ZF subspace, and in most cases also its dimension, is fixed by a large-scale gain ordering rather than optimized against the local SE trade-off. The proposed DAZS framework treats both as joint optimization variables, chosen to maximize the local sum spectral efficiency (SE).

\subsection{Contributions}
The main contributions are summarized as follows.

\begin{itemize}
\item \textbf{Decentralized LSFD under an explicit local information:} We show that, for arbitrary local combining, each AP can construct its  second-layer coefficients using only locally available effective-channel  and interference-plus-noise statistics. The resulting \dLSFD\ is the exact  SINR-maximizing solution for the corresponding diagonal covariance surrogate and  requires no inter-AP statistical exchange. Consequently, the CPU only  aggregates weighted local estimates, avoiding cross-AP covariance acquisition and covariance-matrix inversion for each UE.

\item \textbf{Exact characterization of the decentralization gap:}
We characterize the information discarded by \dLSFD\ through its cross-AP covariance action and establish the exact condition under which \dLSFD\ and \cLSFD\ attain the same SINR. We further introduce the mismatch factor $\Delta_t$ for UE $t$, which decomposes the effect of the omitted covariance into its action strength and directional component. This characterization shows that 
non-negligible cross-AP covariance does not cause a significant SINR loss, particularly when its action on the decentralized decoding direction is weak or predominantly scalar.

\item \textbf{DoF-adaptive ZF-subspace allocation:}
We formulate local ZF as the optimal spatial DoF-allocation problem where both the dimension and composition of the ZF subspace are jointly optimized to maximize the local sum SE. This leads to the \DAZSMR\ and \DAZSPMR\ combining schemes, which combine the adaptively selected ZF subspace with MR and projected-MR (PMR) processing, respectively. Under independent Rayleigh fading, the parallel estimates of pilot-sharing UE enable pilot-domain subspace allocation. Under spatially correlated Rayleigh fading, with generally non-parallel estimates, the same allocation principle is extended to UE directions through a covariance-aware selection surrogate. The resulting subspace-selection problems are solved using continuous relaxation and projected-gradient ascent (PGA).

\item \textbf{Closed-form analysis and performance validation:} For the independent Rayleigh DAZS combining, we derive closed-form SINR expressions and locally computable \dLSFD\ weights. Numerical results show that \dLSFD\ attains within 1\% of the SE of \cLSFD. DAZS combining improves the sum SE by up to 8-11\% over PFZF and PWPFZF combining, while the PGA allocation closely approaches exhaustive-search subspace selection.

\end{itemize}

\subsection{Organization and Notation}
Section~\ref{sec:system} presents the system model. Section~\ref{sec:dlsfd} derives the local-information rule and covariance-action characterization of \dLSFD. Section~\ref{sec:adaptive} develops the DAZS combining framework. Section~\ref{simulations} presents the numerical evaluation, and Section~\ref{sec:conclusion} concludes the paper.

Scalars, vectors and matrices are denoted as $x$, $\mathbf{x}$, and $\mathbf{X}$, respectively. The transpose and Hermitian transpose are denoted by $(\cdot)^T$ and $(\cdot)^H$, respectively, while the complex conjugate is written as $(\cdot)^*$. The sets of real and complex numbers are denoted by $\mathbb{R}$ and $\mathbb{C}$, respectively. The expectation operator is denoted by $\mathbb{E}[\cdot]$. The size $N$ identity matrix  is denoted by $\mathbf{I}_N$. A complex Gaussian random vector $\mathbf{x}$ with mean $\boldsymbol{\mu}$ and covariance matrix $\mathbf{K}$ is denoted as $\mathbf{x} \sim \mathcal{CN}(\boldsymbol{\mu}, \mathbf{K})$. For a vector $\mathbf{x}$, $[\mathbf{x}]_l$ denotes its $l$th element.

\section{System Model}
\label{sec:system}
We consider $M$ APs, each equipped with $A$ antennas, serving $T$ single-antenna UE. UE $t$ is served by the AP set $\mathcal M_t$, and AP $m$ serves the UE set $\mathcal T_m$. The channel is constant over a coherence block of $L_c$ samples, of which $L_p$ are used for uplink pilots. The channel between AP $m$ and UE $t$ follows a Rayleigh model $\mathbf g_{mt}{\sim}\mathcal{CN}(\mathbf0,\mathbf R_{mt})$, where $\mathbf R_{mt}{\succeq}\mathbf0$ is the spatial covariance matrix.

\subsection{Channel Estimation}
Let $\{\boldsymbol\psi_1,{\ldots},\boldsymbol\psi_{L_p}\}$ be the set of orthogonal pilots, with $\|\boldsymbol\psi_i\|^2{=}1$, and let $i_t$ denote the pilot assigned to UE $t$. Define $\mathcal P_i\triangleq\{k:i_k{=}i\}$. The received pilot matrix at AP $m$ is
\begin{equation}
\mathbf Y_m^{\rm p}
=\sum\nolimits_{k=1}^{T}\sqrt{p_k^pL_p}\,\mathbf g_{mk}\boldsymbol\psi_{i_k}^{H}
+\mathbf N_m,
\end{equation}
where the entries of $\mathbf N_m$ are i.i.d. $\mathcal{CN}(0,1)$. After correlating with pilot $i$, the received signal component for pilot $i$ is
\begin{equation}
\mathbf y_{mi}^{\rm p}
=\mathbf Y_m^{\rm p}\boldsymbol\psi_i
=\sum\nolimits_{k\in\mathcal P_i}\sqrt{p_k^pL_p}\,\mathbf g_{mk}
+\mathbf n^p_{mi}.
\end{equation}
where $\mathbf n^p_{mi} {\sim} \mathcal{CN}(0,1)$. Define $\mathbf Q_{mi} {=}\sum\nolimits_{k{\in}\mathcal P_i}p_k^pL_p\mathbf R_{mk}{+}\mathbf I_A$. The MMSE estimate of $\mathbf g_{mt}$ and its estimated covariance are:\begin{align}
\widehat{\mathbf g}_{mt}
&=\sqrt{p_t^pL_p}\,\mathbf R_{mt}\mathbf Q_{mi_t}^{-1}
\mathbf y_{mi_t}^{\rm p},
\label{eq:mmse_general}\\
\mathbf\Phi_{mt}
&\triangleq\mathbb E\{\widehat{\mathbf g}_{mt}\widehat{\mathbf g}_{mt}^{H}\}
=p_t^pL_p\mathbf R_{mt}\mathbf Q_{mi_t}^{-1}\mathbf R_{mt}.
\label{eq:phi_general}
\end{align}
The estimation error $\widetilde{\mathbf g}_{mt}{=}\mathbf g_{mt}-\widehat{\mathbf g}_{mt}$ is independent of $\widehat{\mathbf g}_{mt}$ and has covariance $\mathbf C_{mt}{=}\mathbf R_{mt}-\mathbf\Phi_{mt}$. 

\subsection{Uplink Data Detection}
The received uplink signal at AP $m$ is
\begin{equation}
\mathbf y_m=\sum\nolimits_{k=1}^{T}\sqrt{p_k^u}\,\mathbf g_{mk}x_k+\mathbf n_m,
\end{equation}
where $\mathbb E\{|x_k|^2\}=1$ and $\mathbf n_m\sim\mathcal{CN}(\mathbf0,\mathbf I_A)$. AP $m$ forms the local soft estimate $\mathbf v_{mt}^{H}\mathbf y_m$ and applies a second-layer coefficient $a_{mt}$. The CPU obtains $\widehat x_t= \sum\nolimits_{m\in\mathcal M_t}a_{mt}^{*}\mathbf v_{mt}^{H}\mathbf y_m$. For an arbitrary combiner, define the effective channel vector $\mathbf u_{tk} \triangleq \left[\mathbf v_{mt}^{H}\mathbf g_{mk}\right]_{m\in\mathcal M_t}$ and  its mean $\boldsymbol\mu_{tk}\triangleq\mathbb E\{\mathbf u_{tk}\}$, then
\begin{align}
\mathbf d_t\triangleq \sqrt{p_t^u}\,\boldsymbol\mu_{tt},
\label{eq:d_general} \quad
\mathbf F_t
\triangleq 
\diag\!\left(
\mathbb E\{\|\mathbf v_{mt}\|^2\}
\right)_{m\in\mathcal M_t}.
\end{align}
Using the use-and-then-forget bound~\cite{medard2002effect,marzetta2016fundamentals,bjornson2017massive}, the SE is
\begin{align}
\label{SE_eq}
& \SE_t^u=\frac{L_u}{L_c}\log_2(1+\SINR_t),
\quad
L_u=\frac{L_c-L_p}{2},
\\
&\text{where} \quad  \SINR_t(\mathbf a_t)= 
\frac{|\mathbf a_t^H\mathbf d_t|^2}
{\mathbf a_t^H\mathbf J_t\mathbf a_t},
\label{eq:generic_sinr} \\
& \mathbf J_t
=\sum\nolimits_{k=1}^{T}p_k^u\mathbb E\{\mathbf u_{tk}\mathbf u_{tk}^H\}
-\mathbf d_t\mathbf d_t^H+\mathbf F_t.
\label{eq:J_generic}
\end{align}
The SINR-maximizing centralized LSFD direction is
\begin{equation}
\mathbf a_t^{\mathrm c}\propto\mathbf J_t^{-1}\mathbf d_t.
\label{eq:central_weight}
\end{equation}
For compactness, define $\mu_{mkt}\triangleq \mathbb E\{\mathbf v_{mt}^{H}\mathbf g_{mk}\}$ and
\begin{align}
[\mathbf J_t]_{mn}
=&\sum\nolimits_{k\neq t}p_k^u\mu_{mkt}\mu_{nkt}^{*},
\qquad m\neq n.
\label{eq:J_offdiag}
\end{align}
The off-diagonal LSFD covariance, $[\mathbf J_t]_{mn}$, arises only from coherent statistical coupling among the local soft estimates.

\section{Principled Decentralized LSFD}
\label{sec:dlsfd}
This section develops d-LSFD under an explicit local-information restriction and characterizes the information discarded by decentralization. 

\subsection{Locally Constructible Covariance Surrogate}
Define $ \mathbf D_t\triangleq\diag(\mathbf J_t)$. The corresponding locally constructible surrogate SINR is
\begin{equation}
{\SINR}^{\text{surrogate}}_t(\mathbf a_t)
=\frac{|\mathbf a_t^H\mathbf d_t|^2}
{\mathbf a_t^H\mathbf D_t\mathbf a_t} =\frac{|\mathbf u^H\mathbf q|^2}{\|\mathbf u\|^2}
\le \|\mathbf q\|^2.
\label{eq:surrogate_sinr}
\end{equation}
The equality exists if and only if $\mathbf u\propto\mathbf q$. Hence,
\begin{equation}
\mathbf a_t^{\mathrm d}\propto\mathbf D_t^{-1}\mathbf d_t \triangleq [\mathbf d_t]_m/[\mathbf J_t]_{mm}.
\label{eq:d_weight}
\end{equation}
Thus, $a_{mt}^{\mathrm d}$ is computed locally and the CPU only aggregates weighted soft estimate and does not require cross-AP covariance exchange and per-UE LSFD matrix inversion.

\subsection{Covariance-Action Characterization}
Define the diagonally normalized covariance and normalized desired direction as
\begin{align}
\!\!\!\!\mathbf B_t
{\triangleq}\mathbf D_t^{-1/2}\mathbf J_t\mathbf D_t^{-1/2},
\ \ 
\mathbf z_t
{\triangleq}
\frac{\mathbf D_t^{-1/2}\mathbf d_t}
{\|\mathbf D_t^{-1/2}\mathbf d_t\|}.
\label{eq:z_def}
\end{align}
By construction, $\mathbf B_t$ has unit diagonal and $\sum_mw_{mt}=1$, where $w_{mt}{\triangleq}|z_{mt}|^2$. For the structural characterization below, assume $[\mathbf d_t]_m\neq0$ for every $m\in\mathcal M_t$. Then define the normalized AP-wise covariance correction as
\begin{equation}
\zeta_{mt}
\triangleq
\frac{
[(\mathbf J_t-\mathbf D_t)\mathbf D_t^{-1}\mathbf d_t]_m
}{[\mathbf d_t]_m}
\label{eq:zeta_def}
\end{equation}
and let $\boldsymbol{\Zeta}_t{\triangleq}\diag(\zeta_{mt})_{m{\in}\mathcal M_t}$. These quantities satisfy
\begin{equation}
(\mathbf B_t-\mathbf I)\mathbf z_t
=\boldsymbol{\Zeta}_t\mathbf z_t.
\label{eq:zeta_action}
\end{equation}
Thus, $\zeta_{mt}$ is the normalized aggregate correction imposed at AP $m$ by the cross-AP covariance omitted in d-LSFD. Define the weighted scalar mismatch
\begin{equation}
\Delta_t
\triangleq
\min_{\xi\in\mathbb C}
\| (\boldsymbol{\Zeta}_t-\xi\mathbf I)\mathbf z_t\|^2.
\label{eq:Delta_def}
\end{equation}
Also, define $ \alpha_t \triangleq \mathbf z_t^H\mathbf B_t\mathbf z_t$, $m_t \triangleq
\lambda_{\min}(\mathbf B_t)$, and the covariance-action residual
\begin{equation}
\mathbf r_t
\triangleq
\mathbf B_t\mathbf z_t-\alpha_t\mathbf z_t.
\label{eq:r_def}
\end{equation}
Since $\mathbf B_t\succ\mathbf0$, both $\alpha_t$ and $m_t$ are strictly positive.

\begin{theorem}[Covariance-action characterization]
\label{thm:cov_action}
The weighted scalar mismatch in closed-form is 
\begin{align}
\Delta_t
{=}\!\!\!\sum\limits_{m\in\mathcal M_t}\!\!\!\!\!w_{mt}|\zeta_{mt}-\bar\zeta_t|^2 
{=}\frac{1}{2}\sum\limits_{m,n\in\mathcal M_t} \!\!\!\!\!\!w_{mt}w_{nt}|\zeta_{mt}-\zeta_{nt}|^2,
\label{eq:Delta_pairwise}
\end{align}
where $ \bar\zeta_t {=} \!\sum_{m\in\mathcal M_t}\!\!w_{mt}\zeta_{mt}$ is the minimizer of~\eqref{eq:Delta_def}. Moreover,
\begin{equation}
\Delta_t=\|\mathbf r_t\|^2.
\label{eq:Delta_residual}
\end{equation}
The resulting d-LSFD-to-c-LSFD SINR ratio satisfies
\begin{equation}
\rho_t \triangleq \frac{\SINR_t(\mathbf a_t^{\mathrm d})} {\SINR_t(\mathbf a_t^{\mathrm c})}
\ge
\left(
1+\frac{\Delta_t}{\alpha_t m_t}
\right)^{-1}.
\label{eq:Delta_rho_bound}
\end{equation}
Finally,
\begin{equation}
\Delta_t=0
\iff
\mathbf a_t^{\mathrm d}\propto\mathbf a_t^{\mathrm c}
\iff
\rho_t=1.
\label{eq:Delta_exactness}
\end{equation}
\end{theorem}
\begin{proof}
See Appendix~\ref{app:cov_action_proof}.
\end{proof}

Theorem~\ref{thm:cov_action} shows that non-zero cross-AP covariance does not by itself determine the decentralization loss. What matters is the part of its action that changes the relative decoding direction, quantified by $\Delta_t{=}\|\mathbf r_t\|^2$. If all serving APs experience the same normalized correction, $\zeta_{mt}{=}\bar\zeta_t$, then
$\Delta_t{=}0$ and d-LSFD coincides with c-LSFD. For non-zero mismatch, \eqref{eq:Delta_rho_bound} provides a direct performance guarantee. In particular,
$\Delta_t {\le} \epsilon\alpha_t m_t {\Longrightarrow}  \rho_t {\ge} \frac{1}{1+\epsilon}$.
Thus, a small directional mismatch relative to the local covariance scale $\alpha_t m_t$ guarantees that d-LSFD remains close to c-LSFD. To separate the strength and directionality of the omitted covariance action, define
\begin{equation}
A_t \triangleq \|(\mathbf B_t-\mathbf I)\mathbf z_t\|^2 =\sum\nolimits_{m\in\mathcal M_t}w_{mt}|\zeta_{mt}|^2.
\label{eq:A_action}
\end{equation}
Using~\eqref{eq:Delta_pairwise}, $ A_t=|\bar\zeta_t|^2+\Delta_t$.
For $A_t>0$, define the direction-changing action fraction
\begin{equation}
\phi_t\triangleq{\Delta_t}/{A_t},
\qquad 0\le\phi_t\le1,
\label{eq:phi_def}
\end{equation}
with $\phi_t=0$ when $A_t=0$. Therefore, $ \Delta_t=A_t\phi_t$. The pair $(A_t,\phi_t)$ identifies three physically distinct covariance-action regimes. First, $A_t\ll1$ corresponds to a \emph{weak covariance-action regime}: the omitted cross-AP covariance acts only weakly on the \dLSFD\ direction, keeping $\Delta_t$ small irrespective of the direction-changing fraction. Second, an appreciable $A_t$ together with $\phi_t\ll1$ defines a \emph{scalar-action regime}: the covariance action can be strong, but it is predominantly a common scaling of the \dLSFD\ direction and therefore produces only a small $\Delta_t$. Third, appreciable values of both $A_t$ and $\phi_t$ define a \emph{directional-action regime}. Equation~\eqref{eq:Delta_rho_bound} further shows that the SINR consequence of these regimes depends on the mismatch relative to the covariance scale $\alpha_t m_t$; hence,
small $A_t\phi_t/(\alpha_t m_t)$ provides a sufficient condition for near-centralized second-layer performance.

\section{DoF-Adaptive ZF-Subspace (DAZS) Combining}
\label{sec:adaptive}
At each AP, local combining must divide available $A$ spatial DoFs between interference suppression and coherent array gain: each direction assigned to the local ZF subspace helps suppress the interference but becomes unavailable for contributing to the array gain. Therefore, local combining can be interpreted as the optimal DoF allocation problem over available DoFs. MR assigns all DoFs to array gain and LP-MMSE balances them only implicitly, whereas PFZF and PWPFZF fix the allocation through an LSFC-based strong/weak partition that does not weigh the interference removed against the array gain lost~\cite{zhang2021local}. The proposed DAZS combining instead formulates selective local ZF, i.e., ZF over an adaptively selected subset of directions, as the optimal DoF allocation problem over the dimension and composition of the ZF subspace to maximize the local sum SE. Under independent Rayleigh fading, pilot-sharing MMSE estimates are parallel, so the candidate ZF directions reduce to pilot directions and admit the closed-form analysis developed below. Under spatially correlated Rayleigh fading, these estimates are generally non-parallel. In that scenario, the DAZS combining chooses from the locally served UE directions using a covariance-aware selection surrogate, while the resulting combining vectors are constructed from the instantaneous MMSE channel estimates.

\subsection{DAZS Combining for Independent Rayleigh Fading}
\label{subsec:dazs_iid}
For independent Rayleigh fading,
\begin{align*}
\mathbf R_{mt}
&{=}\beta_{mt}\mathbf I_A,
&
\mathbf Q_{mi}
&{=}\theta_{mi}\mathbf I_A,
&
\theta_{mi}
&{=}\!\!\sum\nolimits_{k\in\mathcal P_i}p_k^pL_p\beta_{mk}+1,
\end{align*}
where $\beta_{mt}$ is the LSFC and the MMSE estimate becomes
\begin{equation}
\widehat{\mathbf g}_{mt}
=
\sqrt{p_t^pL_p}\,
\beta_{mt}\theta_{mi_t}^{-1}
\mathbf y_{mi_t}^{\rm p}.
\label{eq:mmse_iid}
\end{equation}
Its per-antenna variance is
\begin{equation}
\gamma_{mt}
=
p_t^pL_p\beta_{mt}^2\theta_{mi_t}^{-1}.
\label{eq:gamma_iid}
\end{equation}
Hence, all UE sharing pilot contribute only one independent estimated spatial direction at AP $m$, irrespective of the number of UE in that group. Consequently, the independent Rayleigh realization of DAZS combining performs the ZF-subspace allocation directly in the pilot domain. Collect one observation per orthogonal pilot as $\bar{\mathbf G}_m \triangleq [\mathbf y_{m1}^{\rm p},\ldots,\mathbf y_{mL_p}^{\rm p}] = \mathbf Y_m^{\rm p}\boldsymbol\Psi$, where $\boldsymbol\Psi=[\boldsymbol\psi_1,\ldots \boldsymbol\psi_{L_p}]$.
Then
\begin{equation}
\widehat{\mathbf g}_{mt}
=
\sqrt{p_t^pL_p}\,
\beta_{mt}\theta_{mi_t}^{-1}
\bar{\mathbf G}_m\mathbf e_{i_t}.
\end{equation}
Thus, the DAZS combining decision reduces to selecting which pilot directions should span the local ZF subspace. The SINR in~\eqref{eq:generic_sinr} can be expanded as
\begin{align}
&\!\!\!\!\SINR_t
{=}
\frac{
p_t^u|\textsc{DS}_t|^2
}{
\!\!\sum\nolimits_{k=1}^T
p_k^u\mathbb E\!\left\{|\textsc{UI}_{tk}|^2\right\}
{-}p_t^u|\textsc{DS}_t|^2
{+}\mathbb E\!\left\{|\textsc{GN}_t|^2\right\}
},
\label{eq:dap_uatf_expanded}\\
&\!\!\!\!\!\textsc{DS}_t{=}\!\!\sum\nolimits_{m\in\mathcal M_t} \!\!\!\!a_{mt}^*\mathbb E\!\left\{\mathbf v_{mt}^{H}\widehat{\mathbf g}_{mt}\right\},
\ 
\textsc{UI}_{tk}{=}\!\!\sum\nolimits_{m\in\mathcal M_t}\! \!\!a_{mt}^*\mathbf v_{mt}^{H}\mathbf g_{mk},
 \nonumber \\  
&\!\!\!\!\textsc{GN}_t=\sum\nolimits_{m\in\mathcal M_t}
a_{mt}^*\mathbf v_{mt}^{H}\mathbf n_m. \nonumber 
\end{align}
Let $\mathcal S_m$ denote the set of pilot directions selected for ZF and $\mathcal W_m{=}\{1,{\ldots},L_p\}\setminus\mathcal S_m$ the remaining directions. Define
\begin{equation}
\delta_{mi}
=
\begin{cases}
1,& i\in\mathcal S_m,\\
0,& i\in\mathcal W_m,
\end{cases}
\qquad
L_{\mathcal S_m}
=
\sum_{i=1}^{L_p}\delta_{mi}.
\label{eq:delta}
\end{equation}
Feasible local ZF requires $L_{\mathcal S_m}\leq A-1$. Let $\mathbf E_{\mathcal S_m}$ collect the columns of $\mathbf I_{L_p}$ corresponding to the selected pilot directions. For a selected pilot $i$, let $\boldsymbol\varepsilon_{j_{mi}}$ select its position in $\mathbf E_{\mathcal S_m}$. The normalized local ZF vector is
\begin{align}
\mathbf v_{mi}^{\mathrm{LZF}}
&{=}\!\sqrt{\!(\!A{-}L_{\mathcal S_m}\!)\theta_{mi}}\mathbf{\bar{G}}_m\mathbf{E}_{{\mathcal{S}_m}}\!\!\Big(\!\mathbf{E}^{H}_{{\mathcal{S}_m}}\!\!\mathbf{\bar{G}}^{H}_m\mathbf{\bar{G}}_m\mathbf{E}_{{\mathcal{S}_m}}\!\!\Big)^{{-}1}\!\!\!\!\boldsymbol{\varepsilon}_{j_{mi}}.
\label{eq:lzf}
\end{align}
The normalized MR vector is
\begin{equation}
\mathbf v_{mi}^{\mathrm{MR}}
=\frac{\bar{\mathbf G}_m\mathbf e_i}
{\sqrt{A\theta_{mi}}}.
\label{eq:mr}
\end{equation}

\subsection{DAZS-MR Combining}
\label{subsec:dazs_mr}
The \DAZSMR\ combining assigns the LZF vector in~\eqref{eq:lzf} to each ZF selected pilot direction and the MR vector in~\eqref{eq:mr} to each non-ZF selected direction. Thus, the optimized set $\mathcal S_m$ determines both the ZF-subspace dimension and the directions on which the available suppression DoFs are spent. Define
\begin{align}
b_{mkt}^{\DAZSMRmath}
&=
\sqrt{
\left(
A-\delta_{mi_t}L_{\mathcal S_m}
\right)\gamma_{mk}
},
\label{eq:b_dapcmr}
\\
W_{mt}^{\DAZSMRmath}
&=
\sum\nolimits_{k=1}^{T}
p_k^u
\left(
\beta_{mk}
-
\delta_{mi_t}\delta_{mi_k}\gamma_{mk}
\right)
+1.
\label{eq:W_dapcmr}
\end{align}
The achievable SE is obtained from~\eqref{SE_eq} using
\begin{align}
\SINR_t^{\DAZSMRmath}
\!\!=\!\!\frac{p_t^u\Big|\sum\limits_{m\in\mathcal M_t}a^*_{mt}b_{mtt}^{\DAZSMRmath}\Big|^2}{\!\!\!\!\!\!\!\!\!\sum\limits_{k\in\mathcal P_{i_t}\setminus\{t\}}\!\!\!\!\!\!\!\!p_k^u\Big|\!\!\!\!\!\!\!\sum\limits_{\ \ \ m\in\mathcal M_t}\!\!\!\!\!\!\!\!a^*_{mt}b_{mkt}^{\DAZSMRmath}\Big|^2 \!\!\! +\!\!\!\!\!\sum\limits_{m\in\mathcal M_t}\!\!\!\!\!\!|a^*_{mt}|^2W_{mt}^{\DAZSMRmath}}.
\label{eq:sinr_dapcmr}
\end{align}
The corresponding locally computable \dLSFD\ coefficient is
\begin{equation}
a_{mt}^{\mathrm d,\DAZSMRmath}
=
\frac{
\sqrt{p_t^u}\,b_{mtt}^{\DAZSMRmath}
}{
\sum\nolimits_{k\in\mathcal P_{i_t}\setminus\{t\}}
p_k^u
(b_{mkt}^{\DAZSMRmath})^2
+
W_{mt}^{\DAZSMRmath}
}.
\label{eq:d_dapcmr}
\end{equation}
The derivations of the closed-form SINR and \dLSFD\ expressions are given in Appendix~\ref{apx_1}. The product $\delta_{mi_t}\delta_{mi_k}$ in~\eqref{eq:W_dapcmr} captures the target dependent nature of \DAZSMR\ combining. A ZF selected target uses a local ZF vector and therefore suppresses the estimated components of the other selected pilot directions. A non-ZF selected target uses MR and does not inherit these nulling constraints.

\subsection{DAZS-PMR Combining}
\label{subsec:dazs_pmr}

The \DAZSPMR\ combining retains local ZF for the ZF selected pilot directions and uses PMR combining for the remaining directions
\begin{equation}
\mathbf v_{mi}^{\mathrm{PMR}}
=
\frac{
\mathbf B_m\bar{\mathbf G}_m\mathbf e_i
}{
\sqrt{(A-L_{\mathcal S_m})\theta_{mi}}
},
\qquad
i\in\mathcal W_m,
\end{equation}
where $\mathbf B_m{=}\mathbf I_A {-} \bar{\mathbf G}_m\mathbf E_{\mathcal S_m}
\left( \mathbf E_{\mathcal S_m}^{H} \bar{\mathbf G}_m^{H}\bar{\mathbf G}_m
\mathbf E_{\mathcal S_m} \right)^{-1} \mathbf E_{\mathcal S_m}^{H}\bar{\mathbf G}_m^{H}$. The coherent gain and the local residual term are
\begin{align}
b_{mkt}^{\DAZSPMRmath}
&=
\sqrt{(A-L_{\mathcal S_m})\gamma_{mk}},
\label{eq:b_dapcpmr}
\\
W_{mt}^{\DAZSPMRmath}
&=
\sum\nolimits_{k=1}^{T}
p_k^u
\left(
\beta_{mk}
-\delta_{mi_k}\gamma_{mk}
\right)
+1.
\label{eq:W_dapcpmr}
\end{align}
The corresponding SINR is
\begin{align}
\SINR_t^{\DAZSPMRmath}
\!\!=\!\!\frac{p_t^u\Big|\sum\limits_{m\in\mathcal M_t}a^*_{mt}b_{mtt}^{\DAZSPMRmath}\Big|^2}{\!\!\!\!\!\!\!\!\!\!\!\!\!\sum\limits_{k\in\mathcal P_{i_t}{\setminus}\{t\}}\!\!\!\!\!\!\!p_k^u\Big|\!\!\!\!\!\!\!\!\!\sum\limits_{\ \ \ \ m\in\mathcal M_t}\!\!\!\!\!\!\!\!\!\!a^*_{mt}b_{mkt}^{\DAZSPMRmath}\!\Big|^2 \!\!\!\! +\!\!\!\!\!\!\!\!\!\!\sum\limits_{ \ \ \ m\in\mathcal M_t}\!\!\!\!\!\!\!\!\!|a^*_{mt}|^2W_{mt}^{\DAZSPMRmath}},
\label{eq:sinr_dapcpmr}
\end{align}
and the \dLSFD\ coefficient becomes
\begin{align}
a_{mt}^{\mathrm d,\DAZSPMRmath}
=
\frac{
\sqrt{p_t^u}\,b_{mtt}^{\DAZSPMRmath}
}{
\sum\nolimits_{k\in\mathcal P_{i_t}\setminus\{t\}}
p_k^u
(b_{mkt}^{\DAZSPMRmath})^2
+
W_{mt}^{\DAZSPMRmath}
}.
\label{eq:d_dapcpmr}
\end{align}
The derivations are given in Appendix~\ref{apx_2}. Unlike~\eqref{eq:W_dapcmr}, the suppression term in~\eqref{eq:W_dapcpmr} does not contain $\delta_{mi_t}$. A non-ZF selected PMR target is projected onto the orthogonal complement of the entire selected subspace and therefore benefits from suppression of every selected estimated pilot direction.

\subsection{Pilot-Domain ZF-Subspace Allocation}
\label{subsec:dazs_iid_allocation}

Each AP $m$ jointly selects the number and identities of the pilot directions assigned to ZF. The binary variable $\delta_{mi}{=}1$ allocates one spatial DoF to pilot direction $i$, whereas $\delta_{mi}{=}0$ leaves that direction on the MR/PMR branch. AP $m$ determines $\boldsymbol\delta_m {=}[\delta_{m1},{\ldots},\delta_{mL_p}]^T$ from
\begin{subequations}
\label{eq:group_opt}
\begin{align}
\max_{\boldsymbol\delta_m}\quad
&
\sum\nolimits_{t\in\mathcal T_m}
\log_2
\left(
1+\SINR_{mt}(\boldsymbol\delta_m)
\right),
\\
\text{s.t.}\quad
&
\delta_{mi}\in\{0,1\},
\quad
i=1,\ldots,L_p,
\label{eq:group_opt1}
\\
&
\sum\nolimits_{i=1}^{L_p}\delta_{mi}
\leq A-1.
\label{eq:group_opt2}
\end{align}
\end{subequations}
Here, $\SINR_{mt}$ is obtained from~\eqref{eq:sinr_dapcmr} or \eqref{eq:sinr_dapcpmr} by retaining only AP $m$.  Since all quantities entering \eqref{eq:group_opt} are functions of local large-scale statistics, the allocation is updated on the large-scale-fading time scale rather than on every coherence block. For a low-complexity solution, the binary and cardinality constraints are handled using the continuous relaxation $\delta_{mi}\in[0,1]$ and the penalty-continuation objective
\begin{align}
f_m(\boldsymbol\delta_m)
&{=}
\!\!\!\sum\nolimits_{t\in\mathcal T_m}
\!\!\!\log_2(1+\SINR_{mt})
{-}
\chi\lambda_1
\!\!\!\sum\nolimits_{i=1}^{L_p}
[\delta_{mi}-\delta_{mi}^2]_+^2
\nonumber\\
&\quad
-
\chi\lambda_2
\Big[
\sum\nolimits_{i=1}^{L_p}\delta_{mi}-A+1
\Big]_+^2 .
\label{eq:penalty}
\end{align}
The relaxed problem is solved locally at each AP using projected gradient ascent (PGA) with backtracking~\cite{li2015accelerated}. For fixed $\chi$, PGA is applied as
\begin{equation}
\boldsymbol\delta_m^{(r+1)}
=
\Pi_{[0,1]^{L_p}}
\left(
\boldsymbol\delta_m^{(r)}
+
\eta_r\nabla f_m(\boldsymbol\delta_m^{(r)})
\right),
\end{equation}
where $\eta_r$ is selected by backtracking. After convergence of the relaxed PGA, the relaxed solution is converted into a feasible binary candidate. If the rounded point exceeds the spatial DoF budget, only the $A-1$ largest components are retained. The resulting procedure is summarized in Algorithm~\ref{alg_nmAPG}.
\begin{algorithm}[!t]
\caption{Projected Gradient Ascent (PGA) for DAZS Combining}
\label{alg_nmAPG}
\begin{algorithmic}[1]

\FORALL{$m\in\mathcal M$ independently}

\STATE Initialize
$\boldsymbol\delta_m^{(0)}$, $\chi=1$, $\lambda_1=5$, $\lambda_2=10$, $\epsilon=0.005$, $\nu=2$, and $r=1$.

\REPEAT
 \REPEAT
        \STATE Select $\eta_r$ by backtracking and update
        $\boldsymbol\delta_m\leftarrow
        \Pi_{[0,1]^{L_p}}(\boldsymbol\delta_m+\eta_r\nabla f_m(\boldsymbol\delta_m))$.
  \UNTIL{relative change of $f^{(r)}_m$ is at most $\epsilon$}
    
    \STATE $r\leftarrow r+1$ and  $\chi\leftarrow\nu\chi$.

\UNTIL{$\Big|\frac{f(\boldsymbol{{\delta}}^{(r)}_m) - f(\boldsymbol{{\delta}}^{(r-1)}_m)}{f(\boldsymbol{{\delta}}^{(r-1)}_m)} \leq \epsilon   \Big|$}

\STATE Binarize  $
\tilde{\delta}_{mi}
=
\begin{cases}
1, & \delta_{mi}^{(r)}\ge0.5,\\
0, & \delta_{mi}^{(r)}<0.5.
\end{cases}
$

\IF{$\sum_{i=1}^{L_p}\tilde{\delta}_{mi}>A-1$}
    \STATE Retain the $A-1$ largest entries of $\boldsymbol\delta_m^{(r)}$ and set all remaining entries to zero.
\ENDIF
\STATE Set the resulting feasible binary vector as $\boldsymbol\delta_m^\star$.
\ENDFOR
\end{algorithmic}
\end{algorithm}

To express the continuous relaxation compactly, define
\begin{equation}
C_{mt}
\triangleq
\sum\nolimits_{k\in\mathcal P_{i_t}{\setminus}\{t\}}
\!\!p_k^u\gamma_{mk},
\quad
G_{mi}
\triangleq
\sum\nolimits_{k\in\mathcal P_i}
\!\!p_k^u\gamma_{mk}.
\label{eq:iid_C_G}
\end{equation}
For \DAZSMR\ combining, 
\begin{align}
&q_{mt}^{\DAZSMRmath} {=} A{-}\delta_{mi_t} \Big(\! 1{+}\!\sum\nolimits_{i\neq i_t}\delta_{mi}
\!\Big),\ \ 
S_{mt}^{\DAZSMRmath} {=} p_t^u q_{mt}^{\DAZSMRmath}\gamma_{mt},
\nonumber
\\
&I_{mt}^{\DAZSMRmath}
=
q_{mt}^{\DAZSMRmath}C_{mt}
{+}
\!\!\sum_{k=1}^{T}\!p_k^u\beta_{mk}
{-}\delta_{mi_t}
\!\!\sum_{i=1}^{L_p}\!\delta_{mi}G_{mi}
{+}1 .
\label{eq:iid_I_mr_relaxed}
\end{align}
For \DAZSPMR\ combining,
\begin{align}
&q_m^{\DAZSPMRmath} = A-\sum\nolimits_{i=1}^{L_p}\delta_{mi}, \ \
S_{mt}^{\DAZSPMRmath} = p_t^u q_m^{\DAZSPMRmath}\gamma_{mt},
\nonumber
\\
&I_{mt}^{\DAZSPMRmath} = q_m^{\DAZSPMRmath}C_{mt} {+}\!\!\sum_{k=1}^{T}p_k^u\beta_{mk}
{-}\!\!\sum_{i=1}^{L_p}\delta_{mi}G_{mi} {+}1 .
\label{eq:iid_I_pmr_relaxed}
\end{align}
These expressions expose the local DoF trade-off: adding a pilot direction to the ZF subspace suppresses its estimated interference component, but reduces the spatial dimension available for coherent combining. The analytical gradients required by PGA are provided in Appendix~\ref{app:gradient}.

\subsection{DAZS Combining under Correlated Rayleigh Fading}
\label{subsec:dazs_correlated}
Under correlated Rayleigh fading, the channel estimates of pilot-sharing UE are generally non-parallel. Hence, unlike independent Rayleigh fading, a pilot-sharing group cannot be represented by a single spatial direction. The DAZS combining principle is retained, but the local ZF subspace is selected directly over the estimated directions of the served UE.

The remaining question is how to select the dimension and composition of the ZF subspace. Unlike independent Rayleigh fading, the correlated channel estimates do not yield the same tractable closed-form local SINR for direct DoF allocation. We therefore construct a statistical SINR surrogate whose sole purpose is to rank candidate ZF subspaces. The surrogate should capture three effects relevant to this allocation: the spatial DoFs remaining after ZF, the channel power of each interferer in the spatial region relevant to the target UE, and the estimated channel power associated with a selected direction that can be credited to the ZF suppression. The surrogate is used only for subspace allocation; once the subspace is selected, the actual ZF, MR, or PMR combining vectors are constructed from the instantaneous correlated MMSE channel estimates. All reported SE and LSFD quantities are subsequently evaluated for these actual effective channels.

To construct these quantities, we seek a beam-domain representation of the correlated channel statistics that provides scalar power measures analogous to those used under independent Rayleigh fading. To this end, we represent the covariance statistics of all UE in a common unitary $A$-point DFT basis $\mathbf U$. The DFT basis is not assumed to exactly diagonalize the covariance matrices; instead, their diagonal entries in this common basis provide beam-wise channel and estimated channel power profiles. For UE $k$ at AP $m$, define
\begin{align*}
\beta_{mk,a}^{\rm b}
{=}
\left[\mathbf U^{H}\mathbf R_{mk}\mathbf U\right]_{aa},
\ \
\gamma_{mk,a}^{\rm b}
{=}
\left[\mathbf U^{H}\boldsymbol{\Phi}_{mk}\mathbf U\right]_{aa},
\end{align*}
where $a=1,\ldots,A$ and $\boldsymbol{\Phi}_{mk}$ is the exact MMSE estimate covariance in~\eqref{eq:phi_general}. Thus, $\beta_{mk,a}^{\rm b}$ describes the channel power of UE $k$ along common beam $a$, while $\gamma_{mk,a}^{\rm b}$ describes the corresponding power contained in its MMSE channel estimate. The average estimated channel power is
\begin{equation}
\bar{\gamma}_{mk}
=
\frac{1}{A}\operatorname{tr}(\boldsymbol{\Phi}_{mk})
=
\frac{1}{A}\sum\nolimits_{a=1}^{A}\gamma_{mk,a}^{\rm b}.
\label{eq:corr_bar_gamma}
\end{equation}
The relevance of an interfering UE $k$ depends on its spatial overlap with the target UE $t$. We therefore use the estimated-channel power of $t$ to identify the beams that are important for its detection. Define the normalized target beam profile
\begin{equation}
\omega_{mt,a}
=
\frac{\gamma_{mt,a}^{\rm b}}
{\sum\nolimits_{b=1}^{A}\gamma_{mt,b}^{\rm b}},
\label{eq:corr_beam_weight}
\end{equation}
where $\sum_a\omega_{mt,a}=1$. A beam carrying a larger fraction of the estimated power of target $t$ consequently receives a larger weight. Using this target-dependent profile, define
\begin{align}
\!\!\beta_{mtk}^{\rm eff}
{=}\! \sum\nolimits_{a=1}^{A} \!\! \omega_{mt,a}\beta_{mk,a}^{\rm b},
\ \
\gamma_{mtk}^{\rm eff}
{=}\! \sum\nolimits_{a=1}^{A} \!\! \omega_{mt,a}\gamma_{mk,a}^{\rm b}.
\label{eq:corr_gamma_eff}
\end{align}
Here, $\beta_{mtk}^{\rm eff}$ measures the channel power of UE $k$ over the beam region emphasized by target UE $t$, and hence represents its effective interference relevance to $t$. Similarly, $\gamma_{mtk}^{\rm eff}$ measures the estimated channel power of UE $k$ over the same target dependent beam region and quantifies the interference reduction credited to selecting its estimated direction for ZF. Thus, the pair$(\beta_{mtk}^{\rm eff},\gamma_{mtk}^{\rm eff})$ compresses the target--interferer spatial relationship into scalar statistics suitable for subspace allocation. The remaining allocation variables specify which UE directions consume the available ZF DoFs. Let $\boldsymbol\delta_m=[\delta_{m1},\ldots,\delta_{mT}]^T$, where $\delta_{mk}=1$ indicates that the estimated direction of UE $k$ is
included in the local ZF subspace. The feasible allocation satisfies
\begin{align}
\!\!\delta_{mk}{\in}\{0,1\},\ \ 
\delta_{mk}{=}0,\ k{\notin}\mathcal T_m,\ \ 
\sum\nolimits_{k=1}^{T}\delta_{mk}{\le} A-1.
\label{eq:corr_dazs_budget}
\end{align}
We now combine the DoF cost and interference-suppression benefit into the selection surrogate. For \DAZSMR\ combining, an unselected target is processed by MR and does not spend DoFs on ZF. If target $t$ is selected, its combining vector is constructed within the ZF branch and
the complete selected subspace consumes spatial DoFs. Accordingly, define the remaining effective array gain factor $q_{mt}^{\DAZSMRmath}$ and the effective residual interference term
$N_{mt}^{\DAZSMRmath}$ as
\begin{align}
q_{mt}^{\DAZSMRmath}
&=
A-\delta_{mt}
\Big(
1+
\sum\nolimits_{j\neq t}
\delta_{mj}
\Big),
\label{eq:corr_mr_q}
\\
N_{mt}^{\DAZSMRmath}
&=
p_t^u
\left(
\beta_{mtt}^{\rm eff}
-\delta_{mt}\gamma_{mtt}^{\rm eff}
\right)
\nonumber\\
&\quad+
\sum\nolimits_{k\neq t}
p_k^u
\left(
\beta_{mtk}^{\rm eff}
-\delta_{mt}\delta_{mk}
\gamma_{mtk}^{\rm eff}
\right).
\label{eq:corr_mr_N}
\end{align}
The factor $q_{mt}^{\DAZSMRmath}$ therefore equals $A$ on the MR branch and decreases with the dimension of the selected ZF subspace when target $t$ belongs to the ZF branch. In $N_{mt}^{\DAZSMRmath}$, $\beta_{mtk}^{\rm eff}$ represents the effective channel power of interferer $k$, while the $\gamma_{mtk}^{\rm eff}$ term credits the reduction obtained when both $t$ and $k$ are processed within the selected ZF subspace. Combining the remaining array gain and residual interference gives the local selection surrogate
\begin{equation}
\widetilde{\SINR}_{mt}^{\DAZSMRmath}
=
\frac{
p_t^u
q_{mt}^{\DAZSMRmath}\bar{\gamma}_{mt}
}{
1+N_{mt}^{\DAZSMRmath}
}.
\label{eq:corr_mr_surrogate}
\end{equation}
Hence, the surrogate explicitly captures the DAZS trade-off: enlarging the ZF subspace can reduce effective interference through the $N_{mt}^{\DAZSMRmath}$ terms, but simultaneously reduces the available spatial gain through $q_{mt}^{\DAZSMRmath}$.

For \DAZSPMR\ combining, every target is projected onto the orthogonal complement of the selected ZF subspace, irrespective of whether its own direction is selected. Consequently, every target pays the DoF cost of the complete selected subspace, while selecting UE $k$ provides an interference--reduction contribution for every projected target. Accordingly,
\begin{align}
q_m^{\DAZSPMRmath}
&=
A-\sum\nolimits_{k\in\mathcal T_m}\delta_{mk},
\label{eq:corr_pmr_q}
\\
N_{mt}^{\DAZSPMRmath}
&=
\sum\nolimits_{k=1}^{T}
p_k^u
\left(
\beta_{mtk}^{\rm eff}
-\delta_{mk}\gamma_{mtk}^{\rm eff}
\right),
\label{eq:corr_pmr_N}
\\
\widetilde{\SINR}_{mt}^{\DAZSPMRmath}
&=
\frac{
p_t^u
q_m^{\DAZSPMRmath}\bar{\gamma}_{mt}
}{
1+N_{mt}^{\DAZSPMRmath}
}.
\label{eq:corr_pmr_surrogate}
\end{align}
Thus, \DAZSPMR\ combining applies the same DoF trade-off as \DAZSMR\ combining, but the selected subspace affects every target through both the common DoF factor $q_m^{\DAZSPMRmath}$ and the projected interference term $N_{mt}^{\DAZSPMRmath}$.  The quantities in~\eqref{eq:corr_mr_surrogate} and \eqref{eq:corr_pmr_surrogate} are covariance-aware \emph{selection surrogates}, used only to compare candidate local ZF subspaces from long-term statistics; they are not claimed to be achievable correlated-Rayleigh SINRs.

\subsection{Correlated-Fading DoF Allocation}
\label{subsec:corr_dazs_optimization}

The correlated-fading allocation retains the structure of \eqref{eq:group_opt}, but replaces pilot-domain variables by UE-domain variables and the independent Rayleigh local SINRs by the covariance-aware selection surrogates in \eqref{eq:corr_mr_surrogate} and \eqref{eq:corr_pmr_surrogate}. For $\mathcal X\in\{\DAZSMRmath,\DAZSPMRmath\}$, AP $m$ solves
\begin{align}
\max_{\boldsymbol\delta_m}
\quad
\sum\nolimits_{t\in\mathcal T_m}
\log_2
\left(
1+\widetilde{\SINR}_{mt}^{\mathcal X}
(\boldsymbol\delta_m)
\right)
\ \ 
\mathrm{s.t.}\ \
\eqref{eq:corr_dazs_budget}.
\label{eq:corr_group_opt}
\end{align}
We use the same continuous relaxation, penalty continuation, PGA, and backtracking procedure as in Algorithm~\ref{alg_nmAPG}. The only structural change is
\begin{align*}
\underbrace{\text{pilot-direction allocation}}
_{\text{independent Rayleigh}}
\quad\longrightarrow\quad
\underbrace{\text{user-direction allocation}}
_{\text{correlated Rayleigh}} .
\end{align*}
Both correlated surrogates admit closed-form analytical gradients, which are given together with their independent-Rayleigh counterparts in Appendix~\ref{app:gradient}.

\subsection{Convergence and Computational Aspects}
\label{subsec:dazs_convergence}

The DAZS combining framework applies backtracking PGA to a continuously relaxed and penalized local objective. For fixed penalty parameters, the objective is continuously differentiable over the compact relaxed domain, and the resulting PGA converges to a first-order stationary
point of the corresponding relaxed problem~\cite{li2015accelerated,hao2024joint}. This does not imply global optimality of the original binary problem.

Under independent Rayleigh fading, each AP optimizes at most $L_p$ pilot-direction variables, whereas under correlated fading it optimizes the $\mathcal{T}_m$ directions. In either case, the required quantities depend only on long-term channel statistics, so the allocation is updated on the large-scale-fading time scale and does not enter the per-symbol payload-processing path.

\section{Performance Evaluation}
\label{simulations}

\subsection{Simulation Setup}
\label{subsec:simulation_setup}
\begin{table*}[!t]
\centering
\footnotesize
\caption{Covariance-action characterization of d-LSFD under independent and spatially correlated Rayleigh fading.}
\label{tab:structural_validation}
\renewcommand{\arraystretch}{1.08}
\setlength{\tabcolsep}{3.0pt}
\begin{tabular}{lcccccccc}
\toprule
&
\multicolumn{4}{c}{\textbf{Independent Rayleigh}}
&
\multicolumn{4}{c}{\textbf{Correlated Rayleigh}}
\\
\cmidrule(lr){2-5}
\cmidrule(lr){6-9}
\textbf{Combiner}
& $\operatorname{med}(A_t)$
& $\operatorname{med}(\phi_t)$
& $\operatorname{med}(\Delta_t)$
& $\mathbb E\{\rho_t\}$
& $\operatorname{med}(A_t)$
& $\operatorname{med}(\phi_t)$
& $\operatorname{med}(\Delta_t)$
& $\mathbb E\{\rho_t\}$
\\
\midrule

MR
& $6.81{\times}10^{-3}$ & 0.957
& $6.33{\times}10^{-3}$ & 0.9883
& $1.38{\times}10^{-3}$ & 0.904
& $1.19{\times}10^{-3}$ & 0.9961
\\

PFZF
& $3.76{\times}10^{-3}$ & 0.974
& $3.54{\times}10^{-3}$ & 0.9901
& $5.07{\times}10^{-4}$ & 0.953
& $4.57{\times}10^{-4}$ & 0.9980
\\

PWPFZF
& $5.37{\times}10^{-3}$ & 0.968
& $5.05{\times}10^{-3}$ & 0.9884
& $6.79{\times}10^{-5}$ & 0.887
& $5.56{\times}10^{-5}$ & 0.9997
\\

\DAZSMR
& $6.93{\times}10^{-3}$ & 0.972
& $6.54{\times}10^{-3}$ & 0.9859
& $1.13{\times}10^{-3}$ & 0.939
& $1.01{\times}10^{-3}$ & 0.9964
\\

\DAZSPMR
& $1.16{\times}10^{-2}$ & 0.964
& $1.09{\times}10^{-2}$ & 0.9815
& $8.65{\times}10^{-4}$ & 0.901
& $7.35{\times}10^{-4}$ & 0.9977
\\

LP-MMSE
& $1.52{\times}10^{-2}$ & 0.963
& $1.42{\times}10^{-2}$ & 0.9774
& $1.56{\times}10^{-3}$ & 0.911
& $1.35{\times}10^{-3}$ & 0.9962
\\

\bottomrule
\end{tabular}
\end{table*}

We consider a CF-mMIMO network with $M{=}100$ APs, each equipped with $A{=}8$ antennas, and $T{=}100$ UE uniformly distributed over a $1{\times}1$~km$^2$ area. A wrap-around topology is used to suppress boundary effects~\cite{bjornson2020scalable}. The LSFCs are generated according to the model in~\cite{ngo2017cell}, including log-normal shadow fading with an $8$~dB standard deviation. Unless otherwise stated, the coherence block contains $L_c{=}200$ channel uses, $L_p{=}7$ orthogonal pilots are available, all UE transmit with $100$~mW, the carrier frequency is $1.9$~GHz, and the bandwidth is $20$~MHz. Each AP serves its strongest UE until at least $95\%$ of its received large-scale channel power is represented, subject to a maximum of $20$ served UE per AP. Every UE is guaranteed at least its strongest AP. The EEM pilot assignment in~\cite{khan2025pilot} is then applied to the resulting network. PFZF and PWPFZF use the conventional cumulative-LSFC strong set rule~\cite{zhang2021local}, subject to the physical ZF budget of at most $A{-}1$ independent directions.  The h-LSFD benchmark uses the LSFC-based weights of~\cite{schulz2024scalable}.

For correlated Rayleigh fading, each AP employs a half-wavelength-spaced uniform linear array (ULA) and the Gaussian local-scattering covariance model~\cite{bjornson2017massive}; the default angular standard deviation is $\sigma_\varphi{=}20^\circ$. For correlated-fading, DAZS combining implementation SE is always evaluated using the actual correlated channel estimates and the corresponding instantaneous local ZF, MR, or PMR combining vectors. Thus, no closed-form SE is used for the correlated-fading results. The results are averaged over $500$ independent network setups.

\subsection{d-LSFD Performance and Characterization}
\label{subsec:dlsfd_results}

\begin{figure}[!h]
\centering
\subfloat[Independent Rayleigh fading.]{
    \includegraphics[width=0.48\textwidth]
    {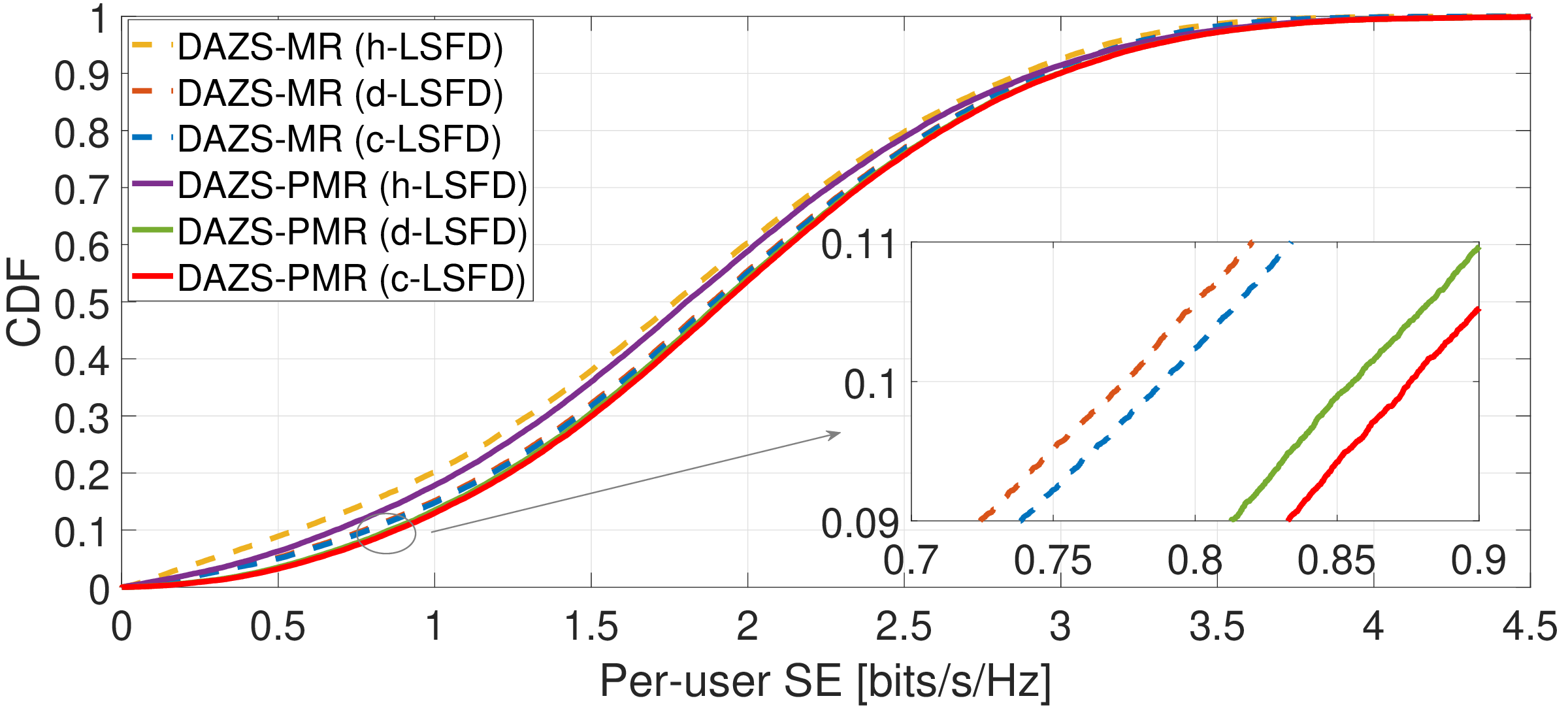}
    \label{fig:decoder_cdf_iid}
}
\vfill
\subfloat[Spatially correlated Rayleigh fading.]{
    \includegraphics[width=0.48\textwidth]
    {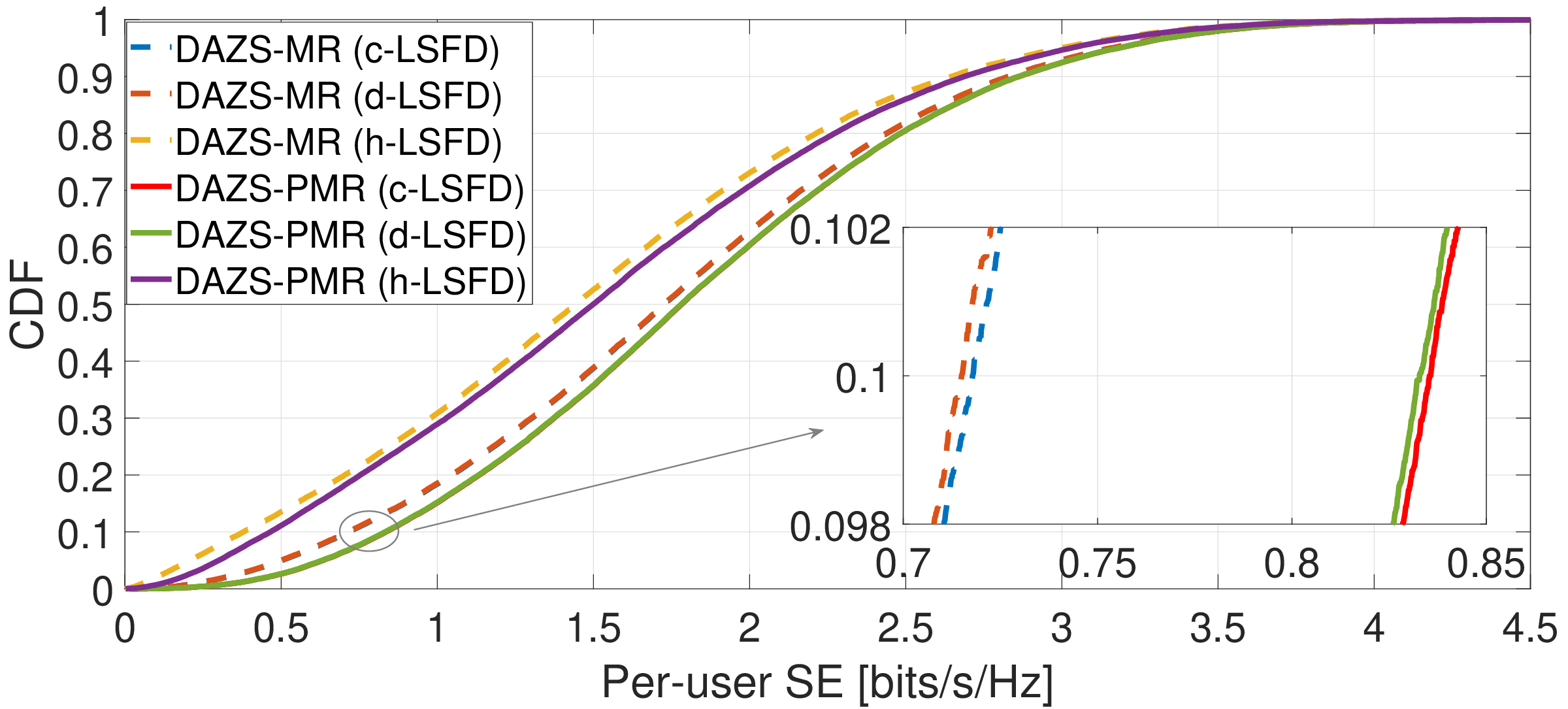}
    \label{fig:decoder_cdf_cor}
}
\caption{CDF of the uplink per-user SE with different decoders.}
\label{fig:decoder_cdf}
\end{figure}
We first evaluate the impact of decentralizing the second-layer combining. Fig.~\ref{fig:decoder_cdf} compares \cLSFD, \dLSFD, and \hLSFD\ for the two proposed local combiners under independent and spatially correlated Rayleigh fading. The \dLSFD\ curves in Fig.~\ref{fig:decoder_cdf} remain close to their \cLSFD\ counterparts for both DAZS combining variants, whereas the LSFC-based \hLSFD\ rule exhibits a substantial deviation. Under independent Rayleigh fading, the $90\%$-likely SE gap between \cLSFD\ and \dLSFD\ is $1.7\%$ and $2.2\%$ for \DAZSMR\ and \DAZSPMR\ combining, respectively. Under correlated fading, the corresponding values are $0.4\%$ and $0.35\%$, respectively.

\begin{table}[!t]
\centering
\footnotesize
\caption{Sum SE under independent and correlated Rayleigh fading.}
\label{tab:overall_SE}
\renewcommand{\arraystretch}{1.08}
\setlength{\tabcolsep}{2.7pt}
\begin{tabular}{lcccccc}
\toprule
\textbf{Combiner}
&
\multicolumn{3}{c}{\textbf{Independent Rayleigh}}
&
\multicolumn{3}{c}{\textbf{Correlated Rayleigh}}
\\
\cmidrule(lr){2-4}
\cmidrule(lr){5-7}
& \cLSFD & \dLSFD & \hLSFD 
& \cLSFD & \dLSFD & \hLSFD 
\\
\midrule

MR
& 121.50 & 120.94 & 100.89 
& 109.62 & 109.43 & 75.25 
\\

PFZF
& 173.83 & 173.34 & 156.88 
& 159.91 & 159.82 & 113.45 
\\

PWPFZF
& 179.08 & 178.46 & 167.60 
& 164.48 & 164.47 & 113.67 
\\

\DAZSMR
& 188.54 & 187.78 & 175.46 
& 176.26 & 176.07 & 149.20 
\\

\DAZSPMR
& 193.93 & 192.88 & 181.57 
& 182.68 & 182.55 & 154.88 
\\

LP-MMSE
& 204.75 & 203.42 & 198.27 
& 207.57 & 207.35 & 194.15 
\\

\bottomrule
\end{tabular}
\end{table}
The covariance-action quantities developed in Section~\ref{sec:dlsfd} explain this near-centralized performance. Table~\ref{tab:structural_validation} reports the action strength
$A_t$, directional fraction $\phi_t$, resulting mismatch $\Delta_t$, and SINR ratio $\rho_t$ for all considered local combiners. A consistent feature of Table~\ref{tab:structural_validation} is that $\phi_t$ is generally large, so the omitted covariance action need not be approximately scalar. Its magnitude $A_t$, however, remains small, which limits $\Delta_t$ and keeps $\rho_t$ close to one. The system therefore operates primarily in the weak-action regime. Under independent Rayleigh fading, the mean SINR ratio ranges from $0.9774$ to $0.9901$ across the six combiners, while under correlated fading it exceeds $0.996$ for every considered combiner. Thus, the small \dLSFD--\cLSFD\ separation is primarily explained by the weak magnitude of the omitted covariance action, rather than by a
predominantly scalar action.

Table~\ref{tab:overall_SE} further shows that \dLSFD\ retains almost all of the \cLSFD\ sum SE, as the gap remains below $1\%$ for all the combiners. For the proposed receivers, the \dLSFD\ loss relative to \cLSFD\ is only $0.41\%$ and $0.54\%$ for \DAZSMR\ and \DAZSPMR\ combining under independent fading, and only $0.11\%$ and $0.07\%$ under correlated fading. In contrast, \hLSFD\ incurs a much larger loss. Hence, the near-centralized behavior of \dLSFD\ persists across both channel models and all evaluated local combiners.

Partial LSFD lies between c-LSFD and d-LSFD. For visual clarity, we omit p-LSFD from the performance plots.

\subsection{Adaptive ZF-Subspace Allocation and SE Performance}
\label{subsec:dazs_results}
\begin{figure}[!h]
\centering
\subfloat[Independent Rayleigh fading.]{
    \includegraphics[width=0.48\textwidth]
    {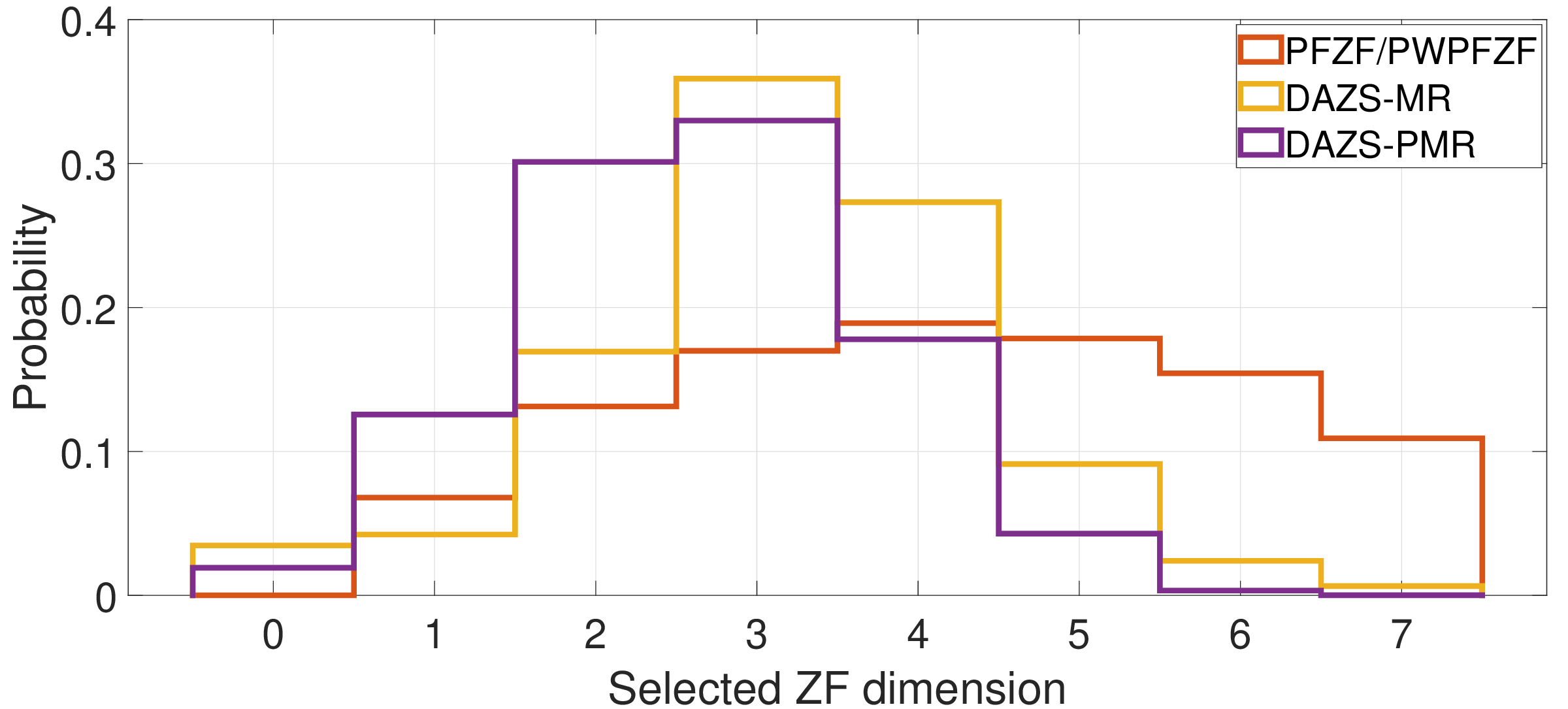}
    \label{fig:zf_hist_iid}
}
\vfill
\subfloat[Spatially correlated Rayleigh fading.]{
    \includegraphics[width=0.48\textwidth]
    {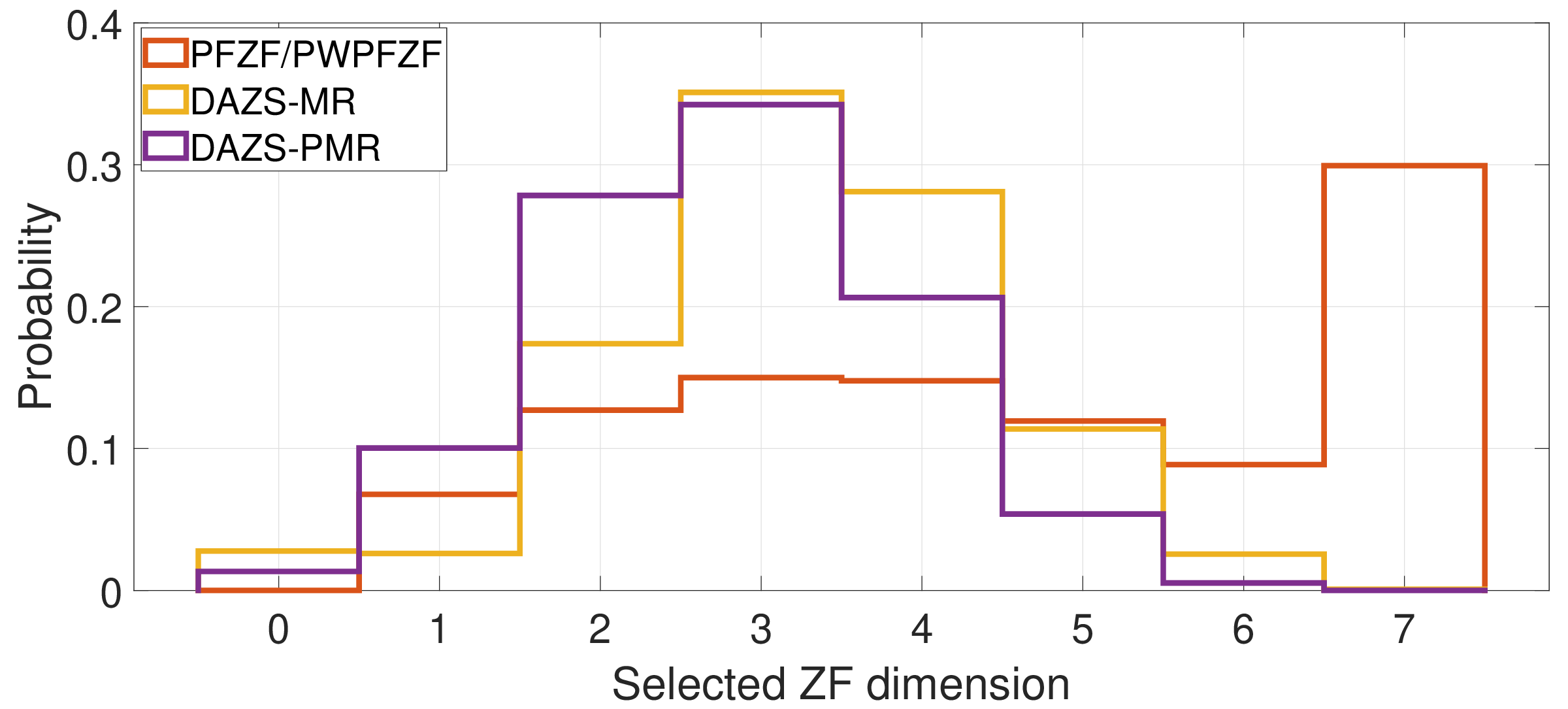}
    \label{fig:zf_hist_cor}
}
\caption{Empirical distribution of the selected local ZF-subspace dimension.}
\label{fig:zf_hist}
\end{figure}
We next examine how DAZS combining allocates the local ZF subspace. Fig.~\ref{fig:zf_hist} shows a clear shift toward smaller ZF dimensions relative to threshold-based grouping. Under independent Rayleigh fading, the mean ZF dimension decreases from $4.18$ for the threshold-based schemes to $3.20$ and $2.66$ for \DAZSMR\ and \DAZSPMR\ combining, respectively; under correlated fading, the corresponding values are $4.59$, $3.28$, and $2.81$. Importantly, DAZS combining assigns zero ZF subspace with non-zero probability, allowing an AP to retain all DoFs for coherent combining when interference suppression is not beneficial. \textbf{Hence, DAZS combining does not merely reduce the ZF dimension; it adapts the amount of interference suppression to the local interference conditions.}

\begin{figure}[!h]
\centering
\subfloat[Independent Rayleigh fading.]{
    \includegraphics[width=0.48\textwidth]
    {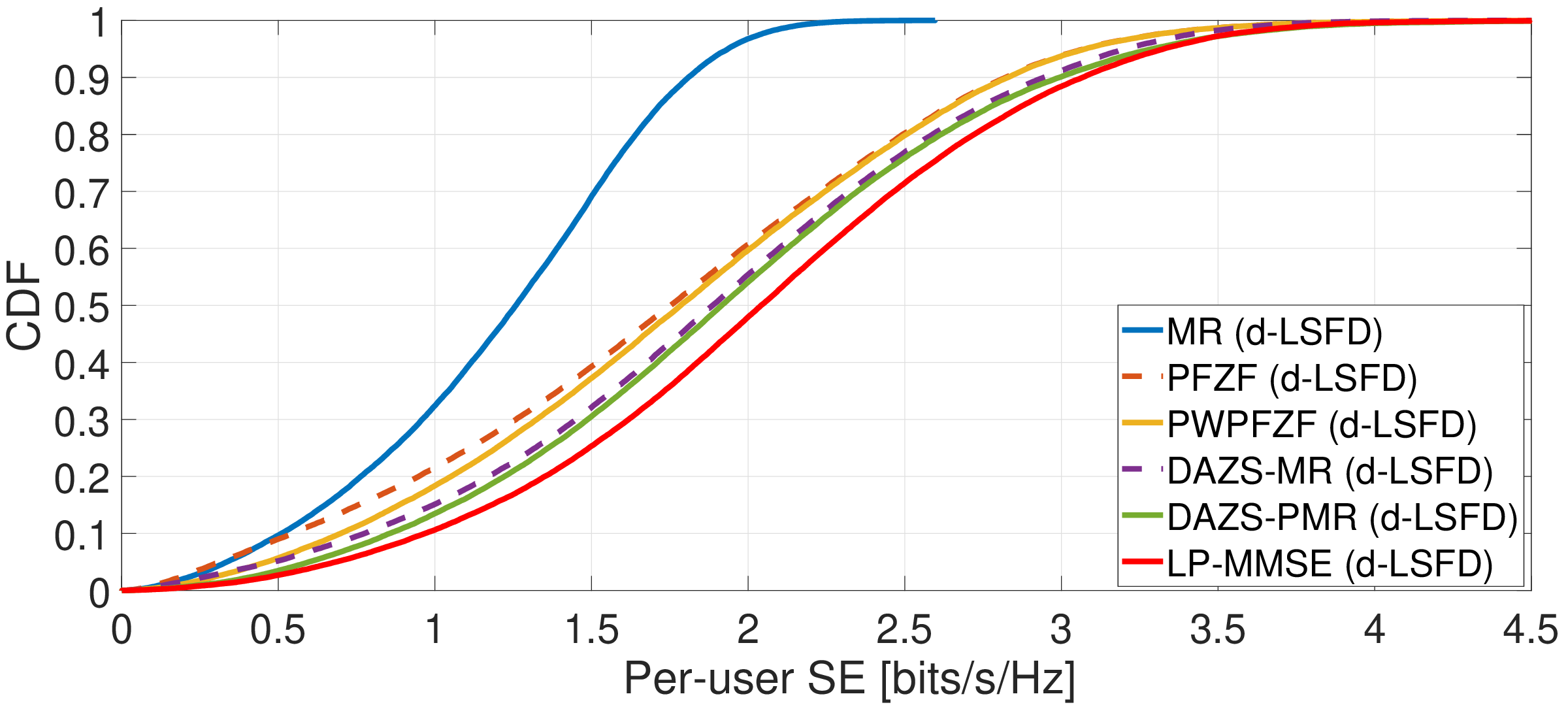}
    \label{fig:allcdf_iid}
}
\vfill
\subfloat[Spatially correlated Rayleigh fading.]{
    \includegraphics[width=0.48\textwidth]
    {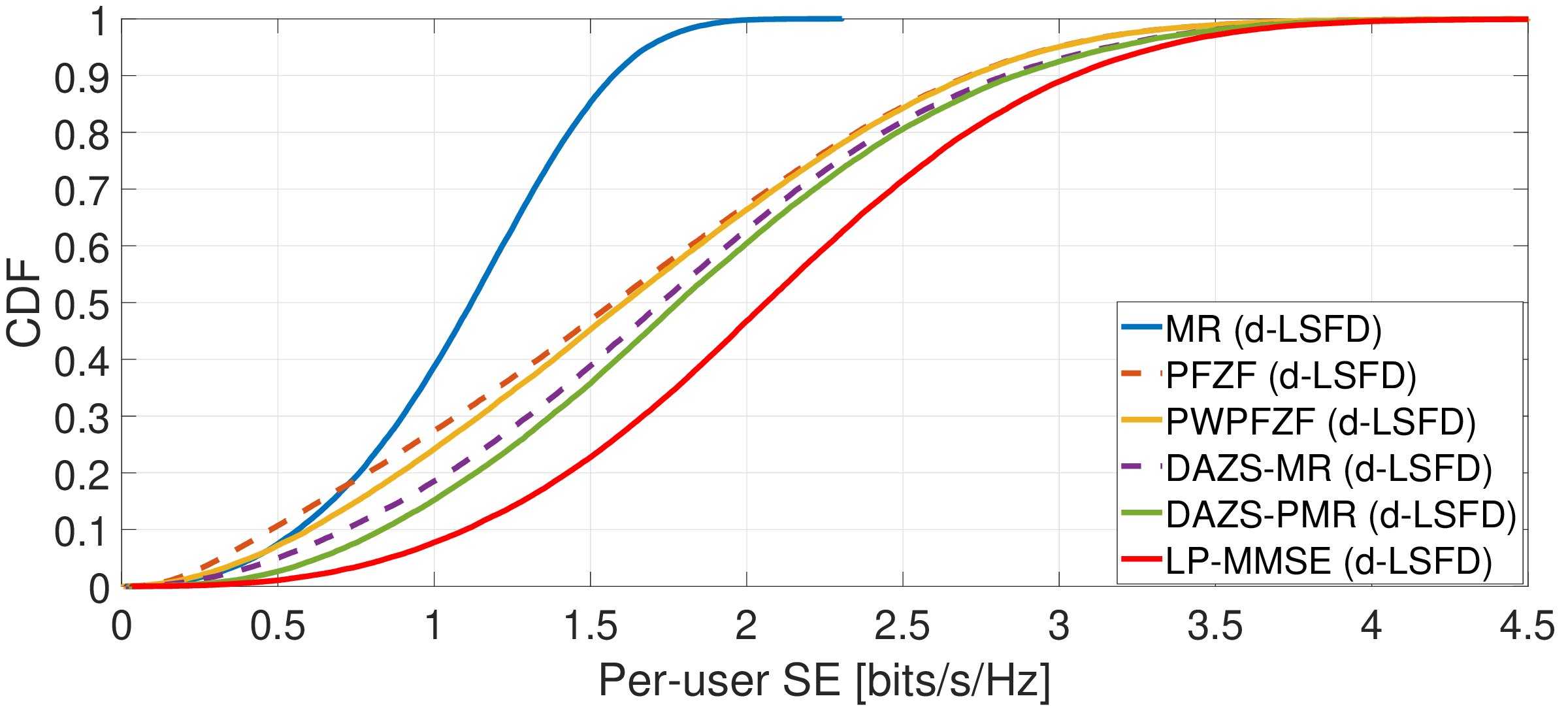}
    \label{fig:allcdf_cor}
}
\caption{CDF of the uplink per-user SE with \dLSFD\ scheme.}
\label{fig:allcdf}
\end{figure}
Fig.~\ref{fig:allcdf} shows that under independent Rayleigh fading, \DAZSMR\ combining increases the $90\%$-likely SE from $0.54$ to $0.77$~bits/s/Hz relative to PFZF, corresponding to a $42\%$ improvement. Likewise, \DAZSPMR\ combining increases it from $0.70$ to $0.86$~bits/s/Hz relative to PWPFZF, yielding a $22\%$ gain. Hence, the benefit of adaptive ZF-subspace allocation extends well into the lower tail of the user-SE distribution rather than being confined to high SE UE. The same behavior is observed under spatially correlated Rayleigh fading. The $90\%$-likely SE increases from $0.47$ to $0.72$~bits/s/Hz for PFZF-to-\DAZSMR\ combining and from $0.60$ to $0.83$~bits/s/Hz for PWPFZF-to-\DAZSPMR\ combining, corresponding to gains of $53\%$ and $39\%$, respectively. LP-MMSE attains the highest overall per-user SE, while the proposed DAZS schemes provide substantial gains over their threshold-based ZF counterparts through adaptive allocation of the local ZF subspace.

\subsubsection{Validation of the DAZS combining optimizer}
\begin{table}[!h]
\centering
\footnotesize
\caption{PGA validation against exhaustive ZF-subspace search.}
\label{tab:pga_exhaustive}
\renewcommand{\arraystretch}{1.08}
\setlength{\tabcolsep}{3.0pt}
\begin{tabular}{lcc}
\toprule
\textbf{Configuration}
& \textbf{Mean gap [\%]}
& \textbf{Maximum gap [\%]}\\
\midrule
IID \DAZSMR, \cLSFD  & 0.6113 & 2.4549\\
IID \DAZSMR, \dLSFD  & 0.5981 & 2.4552\\
IID \DAZSPMR, \cLSFD & 0.0041 & 0.1753\\
IID \DAZSPMR, \dLSFD & 0.0040 & 0.1755\\
\midrule
Correlated \DAZSMR   & 0.0795 & 3.8119\\
Correlated \DAZSPMR  & $1.0{\times}10^{-4}$ & 0.0614\\
\bottomrule
\end{tabular}
\end{table}
Since the binary ZF-subspace selection problem is combinatorial, we validate the proposed PGA implementation against exhaustive subset search. Table~\ref{tab:pga_exhaustive} shows that the PGA solution is very close to exhaustive search. Under independent Rayleigh fading, the mean objective gap is approximately $0.6\%$ for \DAZSMR\ combining and only $0.004\%$ for \DAZSPMR\ combining; the corresponding maximum gaps are $2.46\%$ and $0.18\%$. In the correlated-fading validation, the mean objective gaps are $0.080\%$ and approximately $10^{-4}\%$ for \DAZSMR\ and \DAZSPMR\ combining, respectively.  These results support using PGA in the full-scale simulations, where exhaustive enumeration is computationally impractical.

\subsection{Robustness to Network Loading and Channel Conditions}
\label{subsec:robustness}
\begin{figure}[!t]
\centering
\subfloat[Independent Rayleigh.]{
    \includegraphics[width=0.48\textwidth]
    {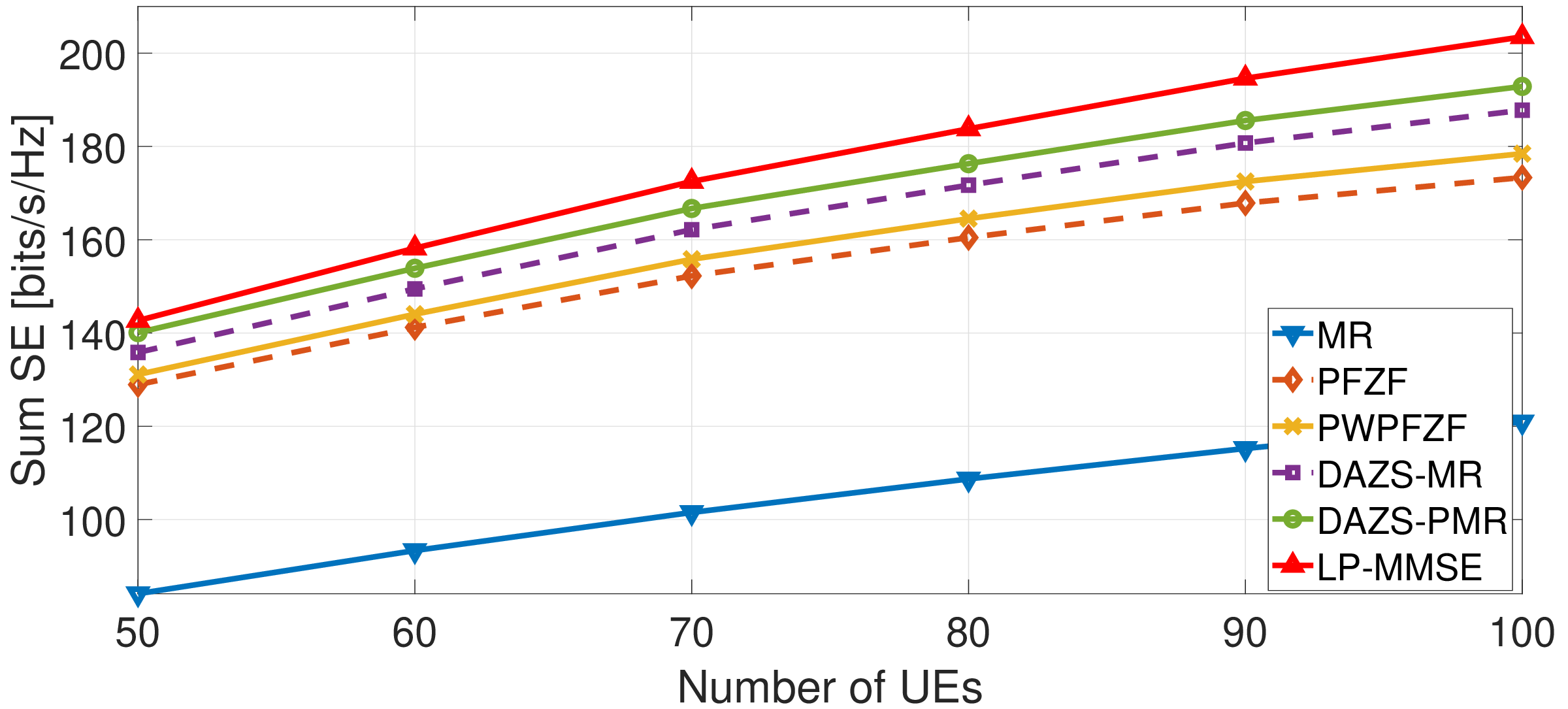}
    \label{fig:user_iid}
}
\hfill
\subfloat[Correlated Rayleigh.]{
    \includegraphics[width=0.48\textwidth]
    {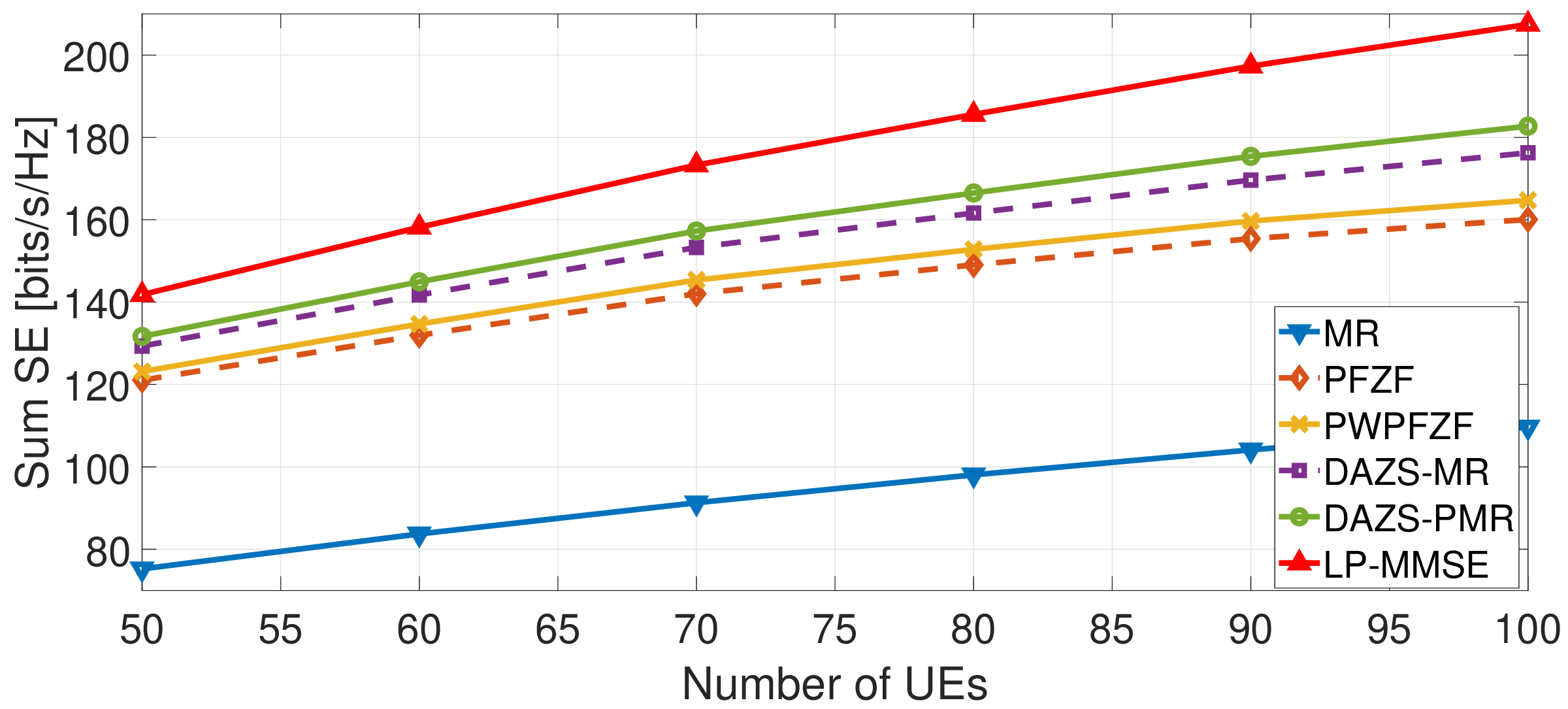}
    \label{fig:user_cor}
}

\subfloat[Independent Rayleigh.]{
    \includegraphics[width=0.48\textwidth]
    {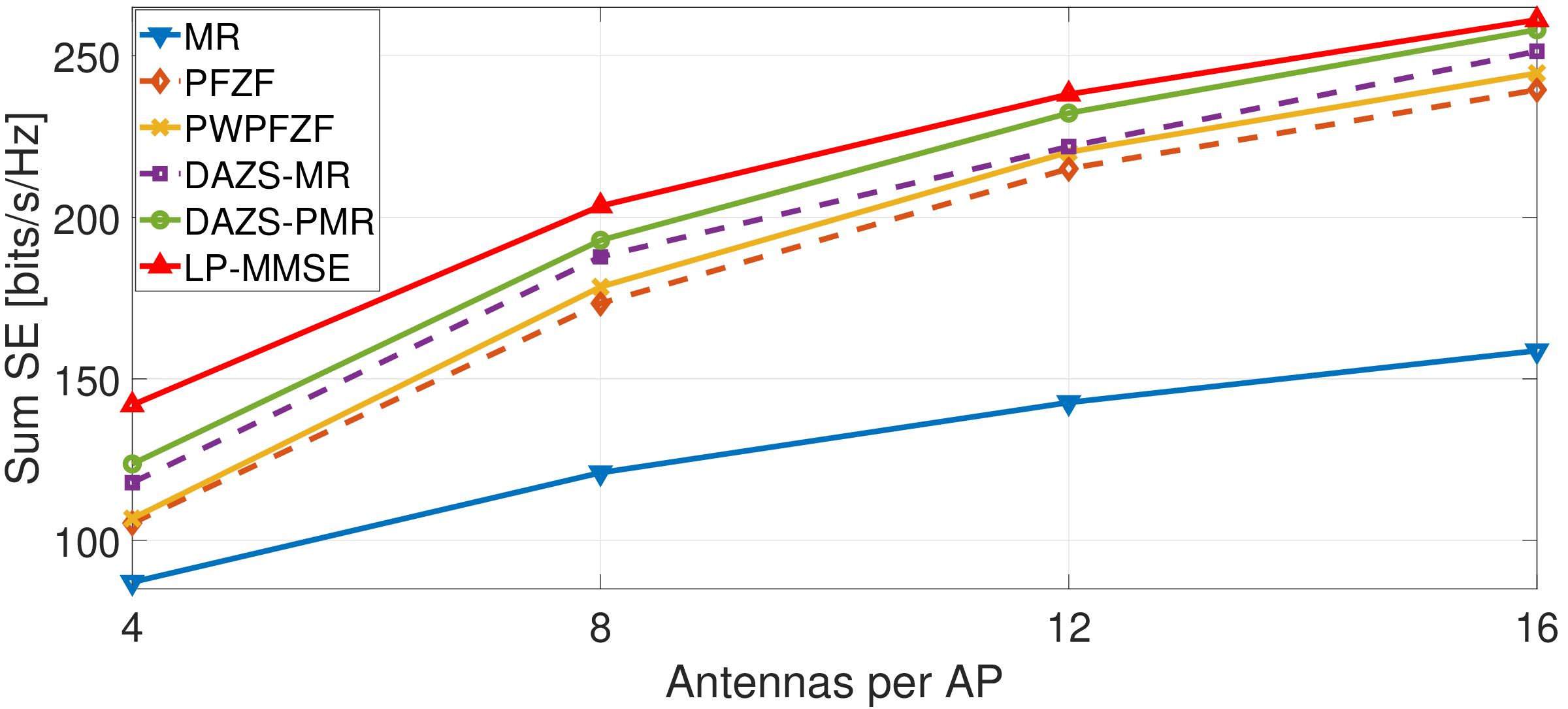}
    \label{fig:ant_iid}
}
\hfill
\subfloat[Correlated Rayleigh.]{
    \includegraphics[width=0.48\textwidth]
    {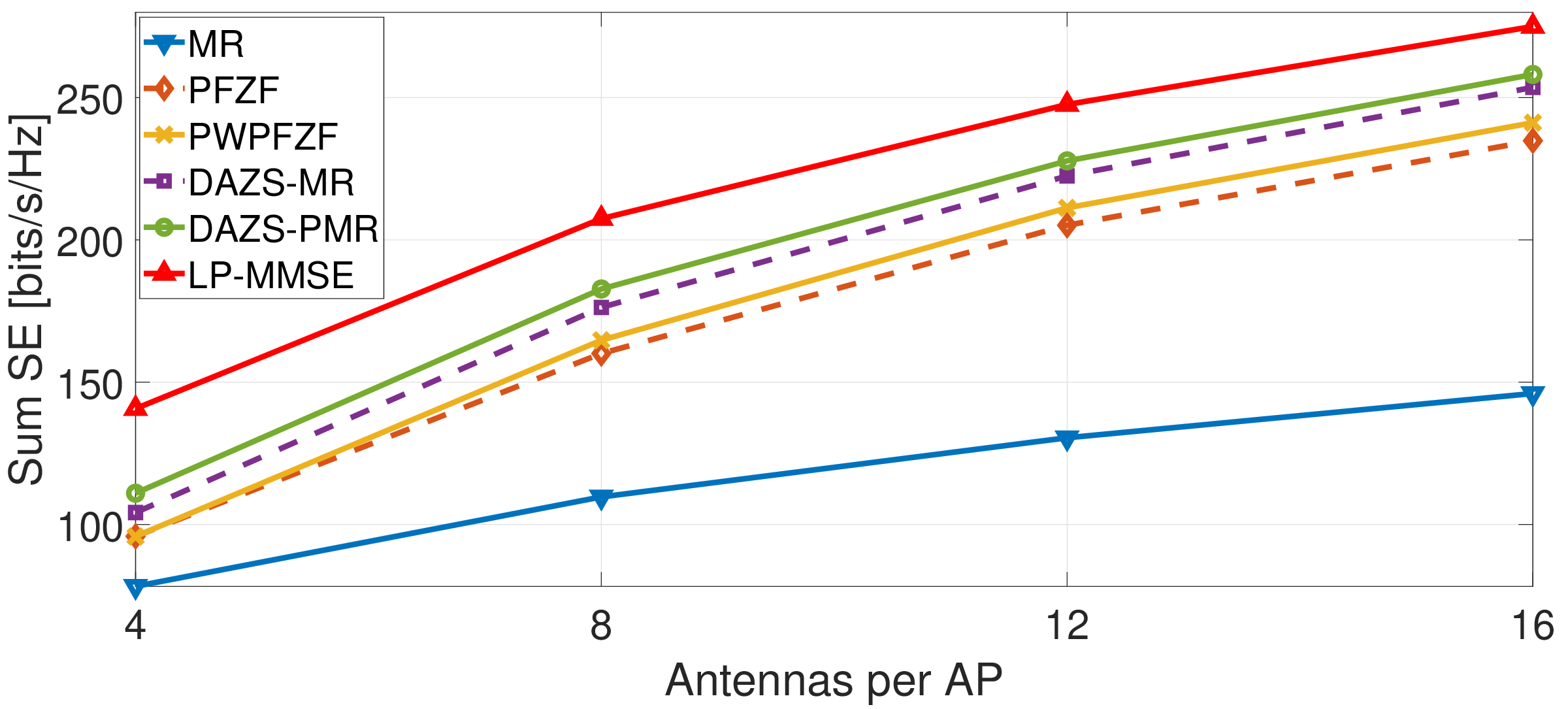}
    \label{fig:ant_cor}
}
\caption{Sum SE with \dLSFD\ versus the number of UE (top) and the number of antennas per AP (bottom).}
\label{fig:load_antenna}
\end{figure}
 We next vary the UE load and the number of antennas while keeping all other parameters unchanged.

Figs.~\ref{fig:user_iid} and~\ref{fig:user_cor} show that the DAZS combining advantage persists over the  range $T{=}50$--$100$ under both channel models. Under independent Rayleigh fading, the sum SE gain of \DAZSMR\ over PFZF changes from 5.3\% at $T{=}50$ to 8.3\% at $T{=}100$, while the corresponding gain of \DAZSPMR\ over PWPFZF changes from 6.8\% to 8\%. Under correlated fading, the respective gains are 6.8\%--10\% and 7\%--11\%. The increasing separation at higher network loads illustrates the value of adaptive spatial-DoF allocation when a limited local ZF budget must be shared under stronger inter-user interference.

Figs.~\ref{fig:ant_iid} and~\ref{fig:ant_cor} examine the impact of the number of antennas per AP over $A\in\{4,8,12,16\}$. Under independent Rayleigh fading, the sum SE gain of \DAZSMR\ over PFZF changes from 12\% at $A=4$ to 5\% at $A=16$, while that of \DAZSPMR\ over PWPFZF changes from 16\% to 5.5\%. Under correlated fading, the corresponding gains are 8.8\%--8\% and 16\%--7\%. The larger relative advantage in the antenna-limited regime reflects the higher cost of an unnecessary ZF constraint when each nulled direction consumes a larger fraction of the available spatial DoFs; as $A$ increases, this cost diminishes, although both DAZS combining variants remain above their threshold-based counterparts throughout the evaluated range.

\begin{figure}[!t]
\centering
\subfloat[Independent Rayleigh.]{
    \includegraphics[width=0.48\textwidth]
    {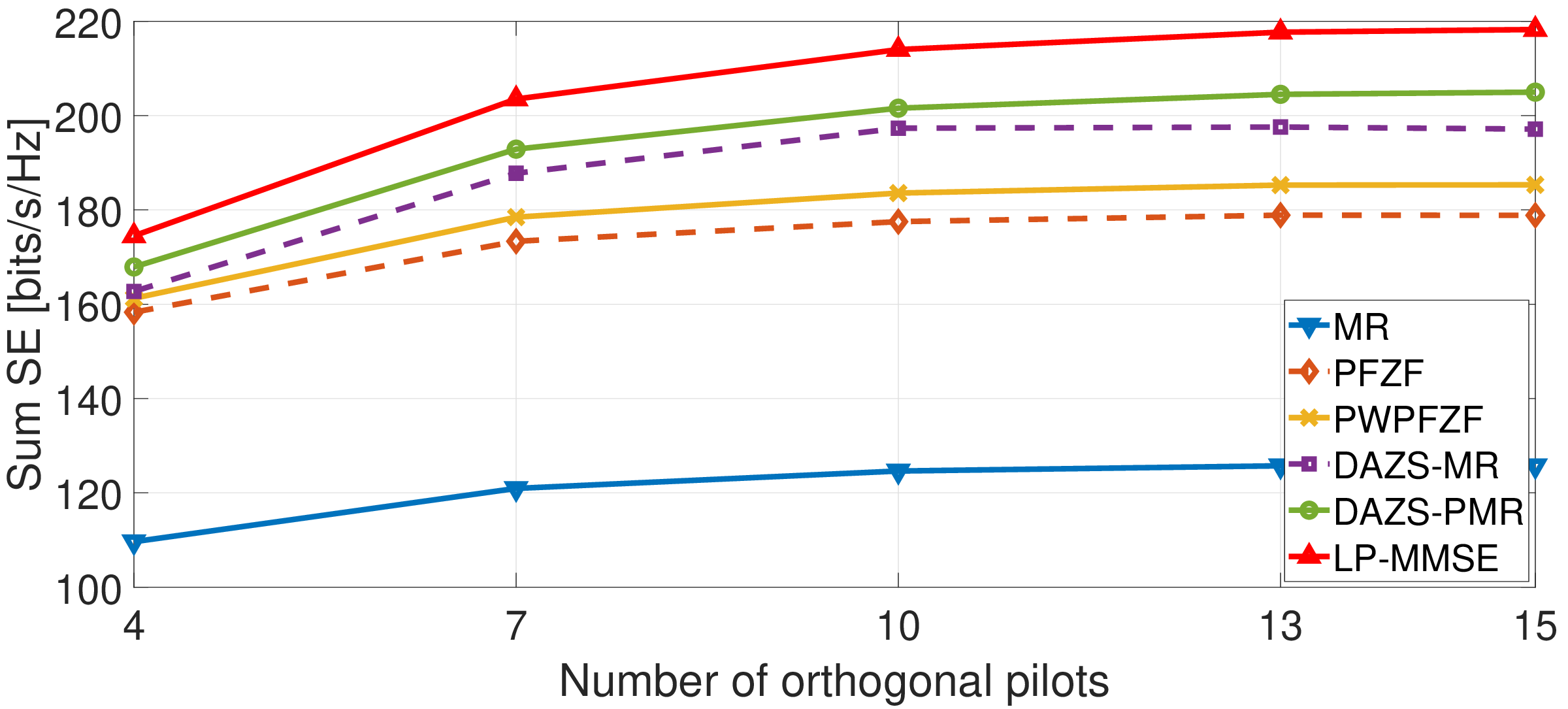}
    \label{fig:pilot_iid}
}
\vfill
\subfloat[Correlated Rayleigh.]{
    \includegraphics[width=0.48\textwidth]
    {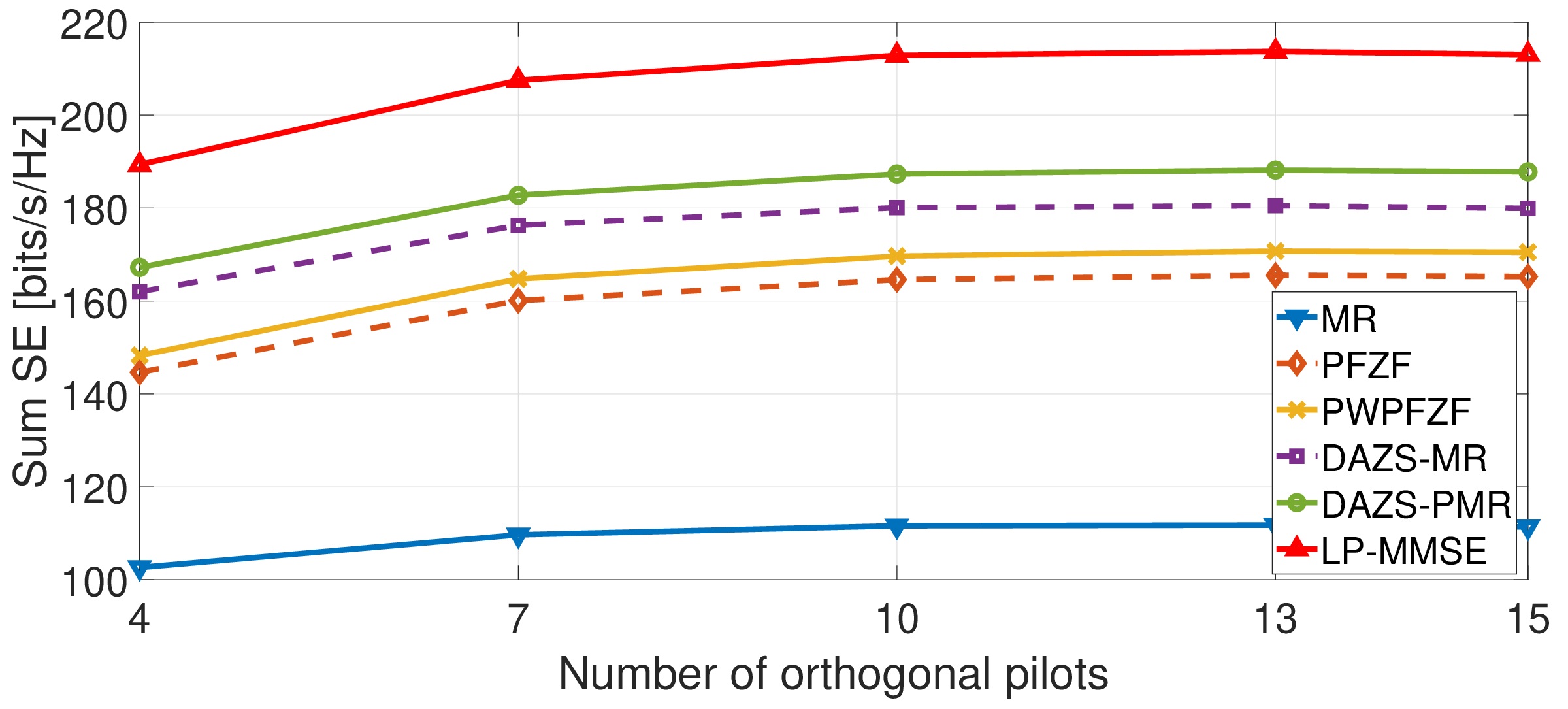}
    \label{fig:pilot_cor}
}
\vfill
\subfloat[Correlated Rayleigh.]{
    \includegraphics[width=0.48\textwidth]
    {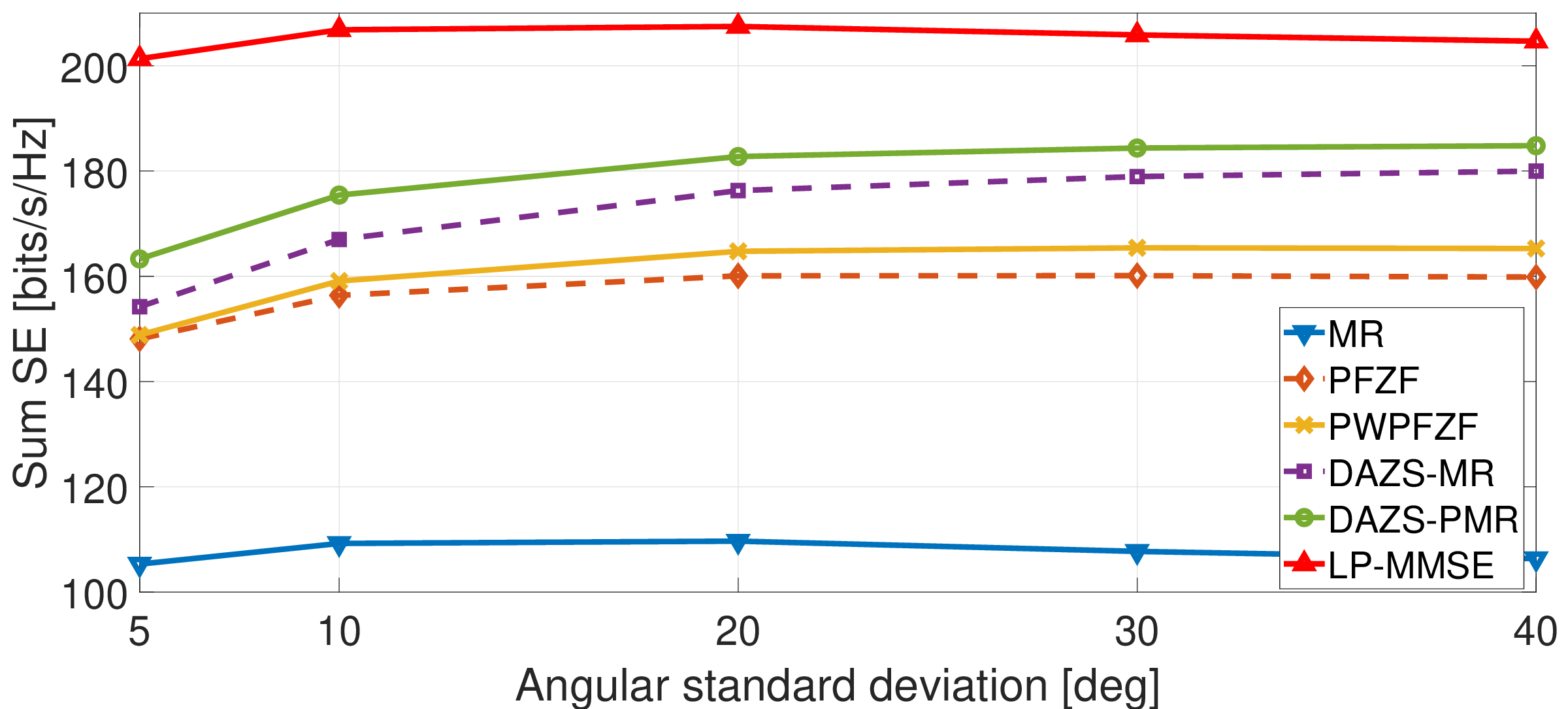}
    \label{fig:asd_cor}
}
\caption{Sum SE with \dLSFD\ versus the pilot length under (a) independent and (b) spatially correlated Rayleigh fading, and (c) versus the angular standard deviation under spatially correlated Rayleigh fading.}
\label{fig:pilot_asd}
\end{figure}
Pilot reuse and spatial correlation are considered in Fig.~\ref{fig:pilot_asd}. Figs.~\ref{fig:pilot_iid} and~\ref{fig:pilot_cor} examine the impact of pilot reuse. Increasing $L_p$ initially improves the sum SE by reducing pilot contamination; beyond this regime, the gain saturates and eventually decreases as the larger pilot overhead reduces the payload fraction of the coherence block. More importantly,  under independent Rayleigh fading, the sum-SE gain of \DAZSMR\ over PFZF ranges from 2.8\% to 10.2\%, while that of \DAZSPMR\ over PWPFZF ranges from 4\% to 10.5\%. Under correlated fading, the corresponding gain ranges are 12\%--9\% and 12.8\%--10\%, respectively. Thus, the advantage of adaptive ZF-subspace allocation persists throughout the considered pilot length range.

Fig.~\ref{fig:asd_cor} examines the sensitivity to spatial correlation by varying the angular standard deviation from $5^\circ$ to $40^\circ$. The advantage of DAZS combining is maintained throughout the entire range: the sum SE gain of \DAZSMR\ over PFZF varies from 4.2\% to 12.6\%, while \DAZSPMR\ improves upon PWPFZF by 9.7\%--11.8\%. Hence, the performance advantage is robust to the degree of spatial correlation and is not specific to the default $20^\circ$ setting.

\subsection{Computational and Fronthaul Scalability}
\label{subsec:complexity_fronthaul}

\begin{table*}[!t]
\footnotesize
\centering
\caption{Second-layer decoding computation and fronthaul overhead.}
\label{tab:lsfd-complexity}
\renewcommand{\arraystretch}{1}
\setlength{\tabcolsep}{2pt}
\begin{tabular}{|l|c|c|c|}
\hline
\textbf{Decoding}
&
\textbf{Payload fronthaul}
&
\textbf{Statistical fronthaul}
&
\textbf{Weight-computation cost}
\\
&
\textbf{per coherence block}
&
\textbf{per LSF update}
&
\textbf{per LSF update}
\\
\hline

\cLSFD~\cite{bjornson2017massive}
&
\(
L_u\displaystyle\sum_{m=1}^{M}|\mathcal T_m|
\)
&
\(2TM\)
&
\(
\begin{aligned}
\displaystyle
&\sum_{t=1}^{T}
|\mathcal M_t|^2
\frac{|\mathcal P_{i_t}|+2}{2}
+\sum_{t=1}^{T}|\mathcal M_t|T+\sum_{t=1}^{T}
\frac{|\mathcal M_t|^3-|\mathcal M_t|}{3}
\end{aligned}
\)
\\
\hline

\pLSFD~\cite{bjornson2017massive}
&
\(
L_u\displaystyle\sum_{m=1}^{M}|\mathcal T_m|
\)
&
\(2TM\)
&
\(
\begin{aligned}
\displaystyle
&\sum_{t=1}^{T}
|\mathcal M_t|^2
\frac{|\mathcal P_{i_t}|+2}{2}
+\sum_{t=1}^{T}|\mathcal M_t||\mathcal K_t|
+\sum_{t=1}^{T}
\frac{|\mathcal M_t|^3-|\mathcal M_t|}{3}
\end{aligned}
\)
\\
\hline

\dLSFD
&
\(
L_u\displaystyle\sum_{m=1}^{M}|\mathcal T_m|
\)
&
\(0\)
&
\(
\displaystyle
\sum_{t=1}^{T}|\mathcal M_t|T
+\sum_{t=1}^{T}|\mathcal M_t|
+\sum_{t=1}^{T}|\mathcal M_t|^2
\)
\\
\hline
\end{tabular}
\end{table*}
We finally compare the implementation costs of the second-layer decoders and local combining schemes in Table~\ref{tab:lsfd-complexity}. Only complex multiplications and divisions are counted, following~\cite{bjornson2017massive,interdonato2020local,chen2020structured,demir2021foundations}. Since payload transmission and LSFD-statistics exchange occur on different time scales, the two fronthaul contributions are reported separately.

The payload term $L_u\sum_m|\mathcal T_m|$ is common to the three architectures: one local soft estimate is generated for every served UE and uplink payload sample. The additional statistical fronthaul is incurred only when the large-scale statistics and the corresponding LSFD weights are updated. Under the adopted accounting, \cLSFD\ and \pLSFD\ require an additional $2TM$ statistical scalars per LSF update. The proposed \dLSFD\ requires no cross-AP statistical exchange for weight construction.

In \dLSFD, AP $m$ constructs its coefficient from its locally available desired-channel and interference-plus-noise statistics and applies the coefficient before forwarding the local soft estimate. The CPU therefore performs only scalar aggregation. In particular, it does not construct a network-wide LSFD covariance matrix and does not perform a per-UE LSFD matrix inversion. The corresponding removal of the centralized inversion term becomes increasingly important as the number of served APs and UE grows.

\section{Conclusion}
\label{sec:conclusion}
This work addresses two coupled limitations of the existing two-stage uplink processing architecture in CF-mMIMO: the cross-AP coordination required by LSFD, and the finite spatial DoFs available for local interference suppression. For arbitrary local combining, \dLSFD\ is the exact SINR-maximizing decoding direction under a diagonal covariance surrogate built entirely from local effective-channel and interference-plus-noise statistics, requiring no cross-AP covariance exchange. The covariance-action analysis establishes the exact condition for \cLSFD--\dLSFD\ equivalence and provides a decomposition of the residual mismatch $\Delta_t$. Numerical results show that the small observed \cLSFD--\dLSFD\ gap stems mainly from weak cross-AP covariance action rather than from an approximately scalar covariance structure.

For local combining, DAZS recasts selective ZF as a spatial DoF allocation in which both the dimension and the composition of the ZF subspace are optimized for local sum SE, rather than fixed by an LSFC threshold. The framework yields \DAZSMR\ and \DAZSPMR\ combining, with pilot-domain allocation under independent Rayleigh fading and UE-domain allocation via a covariance-aware surrogate under correlated fading. In both cases, the allocation is computed from large-scale statistics by a low-dimensional PGA, and instantaneous estimates are used only to construct the combiners. The resulting allocations closely approach exhaustive search and consistently outperform fixed-rule ZF combining in both the sum SE and lower-tail user SE. Overall, the results show that substantial gains can be obtained without centralized LSFD statistics or fixed interference-suppression rules by adapting both cross-AP combining weights and local spatial DoF usage to the available channel statistics.

\appendix

\subsection{Proof of Theorem~\ref{thm:cov_action}}
\label{app:cov_action_proof}
\setcounter{equation}{0}
\renewcommand{\theequation}{A\arabic{equation}}

Let $c_t\triangleq\|\mathbf D_t^{-1/2}\mathbf d_t\|$. From the definitions of $\mathbf B_t$ and $\mathbf z_t$,
\begin{align}
(\mathbf B_t-\mathbf I)\mathbf z_t
&=\frac{1}{c_t}\mathbf D_t^{-1/2}
(\mathbf J_t-\mathbf D_t)\mathbf D_t^{-1}\mathbf d_t.
\end{align}
Using~\eqref{eq:zeta_action}, its $m$th entry is $ [(\mathbf B_t-\mathbf I)\mathbf z_t]_m =\zeta_{mt}z_{mt}$. Then, the scalar-approximation problem in~\eqref{eq:Delta_def} becomes
\begin{equation}
\Delta_t
=\min_{\xi\in\mathbb C}
\sum\nolimits_m w_{mt}|\zeta_{mt}-\xi|^2.
\end{equation}
This weighted least-squares problem is uniquely minimized by $\xi=\bar\zeta_t=\sum_mw_{mt}\zeta_{mt}$. Then
\begin{equation}
\sum_mw_{mt}|\zeta_{mt}-\bar\zeta_t|^2
=\frac{1}{2}\sum_{m,n}w_{mt}w_{nt}|\zeta_{mt}-\zeta_{nt}|^2,
\end{equation}
proving~\eqref{eq:Delta_pairwise}. Next, using~\eqref{eq:zeta_action},
\begin{align}
\alpha_t
{=}
\mathbf z_t^H\mathbf B_t\mathbf z_t
{=}
1+\mathbf z_t^H\boldsymbol{\Zeta}_t\mathbf z_t
{=}
1+\sum_m w_{mt}\zeta_{mt}
{=}
1+\bar\zeta_t.
\label{eq:alpha_zeta_proof}
\end{align}
Hence, the covariance-action residual in~\eqref{eq:r_def} satisfies
\begin{align}
\mathbf r_t = \mathbf B_t\mathbf z_t-\alpha_t\mathbf z_t = (\boldsymbol{\Zeta}_t- \bar\zeta_t\mathbf I)\mathbf z_t.
\label{eq:r_zeta_proof}
\end{align}
 Moreover,
\begin{align*}
\|\mathbf r_t\|^2 = \left\| (\boldsymbol{\Zeta}_t-\bar\zeta_t\mathbf I)\mathbf z_t \right\|^2
= \sum_m w_{mt}|\zeta_{mt}-\bar\zeta_t|^2 = \Delta_t,
\end{align*}
which proves~\eqref{eq:Delta_residual}. We next establish the SINR guarantee.  Let $\mathbf q_t{=}\mathbf D_t^{-1/2}\mathbf d_t{=}c_t\mathbf z_t$. For c-LSFD, $ \SINR_t(\mathbf a_t^{\mathrm c}) {=} \mathbf d_t^H\mathbf J_t^{-1}\mathbf d_t {=}c_t^2\mathbf z_t^H\mathbf B_t^{-1}\mathbf z_t$, whereas d-LSFD yields $
\SINR_t(\mathbf a_t^{\mathrm d}){=} \frac{c_t^2}{\alpha_t}$. Therefore,
\begin{align}
\rho_t
=( \alpha_t \mathbf z_t^H\mathbf B_t^{-1}\mathbf z_t)^{-1}.
\label{eq:rho_cov_action_proof}
\end{align}
Using
$\mathbf B_t\mathbf z_t=\alpha_t\mathbf z_t+\mathbf r_t$
and $\mathbf z_t^H\mathbf r_t=0$, we obtain
\begin{align}
\mathbf r_t^H\mathbf B_t^{-1}\mathbf r_t
&= \alpha_t^2
\mathbf z_t^H\mathbf B_t^{-1}\mathbf z_t
-\alpha_t.
\label{eq:r_inverse_proof}
\end{align}
Since $m_t=\lambda_{\min}(\mathbf B_t)>0$, then $\mathbf B_t^{-1} \preceq \frac{1}{m_t}\mathbf I$, and hence
\begin{equation}
\mathbf r_t^H\mathbf B_t^{-1}\mathbf r_t
\le
\frac{\|\mathbf r_t\|^2}{m_t}
=
\frac{\Delta_t}{m_t}.
\end{equation}
Combining this inequality with~\eqref{eq:r_inverse_proof} yields
\begin{equation}
\rho_t
\ge
\left(
1+\frac{\Delta_t}{\alpha_t m_t}
\right)^{-1},
\end{equation}
which proves~\eqref{eq:Delta_rho_bound}. Finally, from~\eqref{eq:Delta_residual}, $\Delta_t{=}0$ if and only if $\mathbf r_t{=}\mathbf0$, which is equivalent to $ \mathbf B_t\mathbf z_t {=} \alpha_t\mathbf z_t$. Thus, $\Delta_t{=}0$ if and only if $\mathbf z_t$ is an eigenvector of $\mathbf B_t$. Then
\begin{align}
\mathbf a_t^{\mathrm d}
\propto \mathbf D_t^{-1/2}\mathbf z_t, \qquad
\mathbf a_t^{\mathrm c}
\propto \mathbf D_t^{-1/2}\mathbf B_t^{-1}\mathbf z_t.
\end{align}
Hence, $\mathbf a_t^{\mathrm d}\propto\mathbf a_t^{\mathrm c}$ if and only if $\mathbf z_t$ is an eigenvector of $\mathbf B_t$. Since~\eqref{eq:generic_sinr} is a generalized Rayleigh quotient with $\mathbf J_t\succ\mathbf0$, its maximizing direction is unique up to a non-zero scalar. Therefore, $\mathbf a_t^{\mathrm d}\propto\mathbf a_t^{\mathrm c}$ if and only if the two schemes attain the same SINR, i.e., $\rho_t=1$. This proves~\eqref{eq:Delta_exactness}.

\subsection{Closed-Form Expressions for DAZS-MR Combining}
\label{apx_1}
\setcounter{equation}{0}
\renewcommand{\theequation}{B\arabic{equation}}
The local ZF vector in~\eqref{eq:lzf} satisfies
\begin{align}
\mathbb E\!\left\{(\mathbf v_{mi_t}^{\mathrm{LZF}})^H
\widehat{\mathbf g}_{mk}\right\}
&=
\begin{cases}
\sqrt{(A-L_{\mathcal S_m})\gamma_{mk}},&i_k{=}i_t,\\
0,&i_k{\neq} i_t,
\end{cases}
\label{eq_PFZF_cont}\\
\!\!\!\mathbb E\!\left\{\left|(\mathbf v_{mi_t}^{\mathrm{LZF}})^H
\widehat{\mathbf g}_{mk}\right|^2\right\}
&{=}
\begin{cases}
(A-L_{\mathcal S_m})\gamma_{mk},&i_k{=}i_t,\\
0,& \!\!\!\!\!\!\!\!\!\!\!i_k{\neq} i_t,~i_k{\in}\mathcal S_m,\\
\gamma_{mk},& \!\!\!\!\!\!\!\!\!\!\!\!    i_k\in\mathcal W_m,
\end{cases}
\label{eq_PFZF_cont_2}
\end{align}
and $\mathbb E\{\|\mathbf v_{mi_t}^{\mathrm{LZF}}\|^2\}=1$. For a non-ZF-selected target pilot, $i_t\in\mathcal W_m$, normalized MR gives
\begin{align}
\mathbb E\!\left\{(\mathbf v_{mi_t}^{\mathrm{MR}})^H
\widehat{\mathbf g}_{mk}\right\}
&=
\begin{cases}
\sqrt{A\gamma_{mk}},&i_k=i_t,\\
0,&i_k\neq i_t,
\end{cases}
\label{eq_MR_cont}\\
\mathbb E\!\left\{\left|(\mathbf v_{mi_t}^{\mathrm{MR}})^H
\widehat{\mathbf g}_{mk}\right|^2\right\}
&=
\begin{cases}
(A+1)\gamma_{mk},&i_k=i_t,\\
\gamma_{mk},&i_k\neq i_t,
\end{cases}
\label{eq_MR_cont_2}
\end{align}
with $\mathbb E\{\|\mathbf v_{mi_t}^{\mathrm{MR}}\|^2\}=1$.  Define $
\mathcal Z_{i_t}\triangleq\{m\in\mathcal M_t:i_t\in\mathcal S_m\}$ and
$\mathcal Y_{i_t}\triangleq\mathcal M_t\setminus\mathcal Z_{i_t}$. Thus, the target uses LZF at APs in $\mathcal Z_{i_t}$ and MR at APs in $\mathcal Y_{i_t}$. Using~\eqref{eq_PFZF_cont} and~\eqref{eq_MR_cont}, the desired-signal term in~\eqref{eq:dap_uatf_expanded} is
\begin{align}
|\mathrm{DS}_t|^2
&=\!\Big|
\sum_{m\in\mathcal Z_{i_t}}\!\!a_{mt}^*
\sqrt{(A-L_{\mathcal S_m})\gamma_{mt}}
+\sum_{m\in\mathcal Y_{i_t}}\!\!a_{mt}^*
\sqrt{A\gamma_{mt}}
\Big|^2
\nonumber\\
&=\Big|\sum_{m\in\mathcal M_t}a_{mt}^*
 b_{mtt}^{\DAZSMRmath}\Big|^2.
\label{eq_ds_f}
\end{align}

For the interference term, write $\mathbb E\{|\mathrm{UI}_{tk}|^2\}=A_1+A_2+A_3$, where the LZF-only and MR-only contributions, using
$\mathbf g_{mk}=\hat{\mathbf g}_{mk}+\tilde{\mathbf g}_{mk}$ and the independence of the estimation error, are
\begin{align}
A_1
&=\Big|\sum\nolimits_{m\in\mathcal Z_{i_t}}a_{mt}^*
\sqrt{(A-L_{\mathcal S_m})\gamma_{mk}}\Big|^2
\Big|\boldsymbol\psi_{i_t}^H\boldsymbol\psi_{i_k}\Big|^2
\nonumber\\
&+\sum\nolimits_{m\in\mathcal Z_{i_t}}|a_{mt}|^2
\left(\beta_{mk}-\delta_{mi_k}\gamma_{mk}\right),
\label{eq_bu_2_s11}\\
A_2
&=\!\Big|\!\!\sum_{m\in\mathcal Y_{i_t}}\!\!\!\!a_{mt}^*
\sqrt{A\gamma_{mk}}\Big|^2
\Big|\boldsymbol\psi_{i_t}^H\boldsymbol\psi_{i_k}\Big|^2
+\!\!\sum_{m\in\mathcal Y_{i_t}}\!\!|a_{mt}|^2\beta_{mk}.
\label{eq_bu_2_s21}
\end{align}
The cross term is
\begin{align}
A_3
=&2\operatorname{Re}\!\Big\{
\Big(\sum\nolimits_{m\in\mathcal Z_{i_t}}a_{mt}^*
\sqrt{(A-L_{\mathcal S_m})\gamma_{mk}}\Big)
\nonumber\\
&
\Big(\sum\nolimits_{n\in\mathcal Y_{i_t}}a_{nt}^*
\sqrt{A\gamma_{nk}}\Big)^{\!*}
\Big\}
\Big|\boldsymbol\psi_{i_t}^H\boldsymbol\psi_{i_k}\Big|^2.
\label{eq_bu_2_s3}
\end{align}
Combining~\eqref{eq_bu_2_s11}--\eqref{eq_bu_2_s3} yields
\begin{align}
\mathbb E\{|\mathrm{UI}_{tk}|^2\}
\!&=\!\!\Big|\!\!\!\!\sum_{m\in\mathcal M_t}\!\!\!\!a_{mt}^*
\sqrt{\!(\!A{-}\delta_{mi_t}L_{\mathcal S_m}\!)\gamma_{mk}}\Big|^2
\big|\boldsymbol\psi_{i_t}^H\boldsymbol\psi_{i_k}\big|^2
\nonumber\\
&+\sum\nolimits_{m\in\mathcal M_t}\!\!\!|a_{mt}|^2
\!\left(\beta_{mk}{-}\delta_{mi_t}\delta_{mi_k}\gamma_{mk}\right).
\label{eq_pc_f}
\end{align}
Since both local combining vectors are normalized and the receiver noises are independent across APs,
\begin{equation}
\mathbb E\{|\mathrm{GN}_t|^2\}
=\sum\nolimits_{m\in\mathcal M_t}|a_{mt}|^2.
\label{eq_nc_f}
\end{equation}
Substituting~\eqref{eq_ds_f}, \eqref{eq_pc_f}, and~\eqref{eq_nc_f} into~\eqref{eq:dap_uatf_expanded},  gives the SINR in~\eqref{eq:sinr_dapcmr}. Finally, from~\eqref{eq:d_weight}, the \dLSFD\ coefficient uses
\begin{align}
[\mathbf d_t]_m
&=\sqrt{p_t^u}\,b_{mtt}^{\DAZSMRmath},\\
[\mathbf D_t]_{mm}
&=\sum\nolimits_{k\in\mathcal P_{i_t}\setminus\{t\}}p_k^u
\big(b_{mkt}^{\DAZSMRmath}\big)^2+W_{mt}^{\DAZSMRmath}.
\end{align}
These terms give~\eqref{eq:d_dapcmr}.

\subsection{Closed-Form Expressions for DAZS-PMR Combining}
\label{apx_2}
\setcounter{equation}{0}
\renewcommand{\theequation}{C\arabic{equation}}
\begin{align*}
&\mathbb{E}\{(\mathbf{v}_{mi_t}^{\text{PMR}})^{H}{\mathbf{\hat{g}}_{mk}}\} = \begin{cases}
  \sqrt{(A-L_{\mathcal{S}_m}) \gamma_{mk}}& \text{if } i_k =i_t,\\
    0,              & \text{if } i_k \neq i_t.
\end{cases} \\
&\mathbb{E}\Big\{ \Big|(\mathbf{v}_{mi_t}^{\text{PMR}})^{H}\hat{\mathbf{g}}_{mk}\Big|^2\Big\} = \begin{cases}
  (A-L_{\mathcal{S}_m}+1)\gamma_{mk} \ \ \text{if } i_k =i_t,\\
  0,              \ \  \text{if } i_k \neq i_t \ \text{and} \ i_k \in \mathcal{S}_m ,\\
    \gamma_{mk},              \ \ \text{if } i_k \in \mathcal W_m,
\end{cases}
\end{align*}
The derivation follows the same steps as Appendix~\ref{apx_1}, with ${\mathbf{v}}^{\text{PMR}}_{mi_t}$ replacing ${\mathbf{v}}^{\text{MR}}_{mi_t}$ for the non-ZF-selected target. The desired signal term in \eqref{eq:dap_uatf_expanded} 
\begin{align}
\label{eq_ds1_f}
|\text{DS}_t|^2 &= \Big|\mathbb{E}\Big\{\!\!\!\!\! \sum_{m \in \mathcal{Z}_{i_t}}\!\!\!\!\!a^*_{mt}(\mathbf{v}_{mi_t}^{\text{LZF}})^{H}\hat{\mathbf{{g}}}_{mt} +\!\!\!\!\! \sum_{m \in \mathcal{Y}_{i_t}}\!\!\!\!\!a^*_{mt}(\mathbf{v}_{mi_t}^{\text{PMR}})^{H}\hat{\mathbf{{g}}}_{mt} \Big\}\Big|^2  \nonumber \\ &
 = \Big| \sum\nolimits_{m \in \mathcal{M}_t}a^*_{mt}\sqrt{(A-L_{\mathcal{S}_m}) \gamma_{mt}} \Big|^2.
\end{align}  
The first term in \eqref{eq:dap_uatf_expanded} can be written as
\begin{align}
\label{eq_pc1_f}
\begin{split}
\mathbb{E}\Big\{\!|\text{UI}_{tk}|^2 \!\Big\} & =\!\! \Big|\!\! \sum\limits_{ m \in \mathcal{M}_t}\!\!\!a^*_{mt}\sqrt{(A{-}L_{\mathcal{S}_m}) \gamma_{mk}}\Big|^2\!\!\left|\boldsymbol{\psi}^{H}_{i_t}\boldsymbol{\psi}_{i_k} \right|^2  \\ & +   \sum\limits_{m \in \mathcal{M}_t}\!|a^*_{mt}|^2(\beta_{mk} -\delta_{mi_k}\gamma_{mk}).
\end{split}
\end{align} 
The receiver-noise term in~\eqref{eq:dap_uatf_expanded} is
\begin{align}
\label{eq_nc1_f}
\begin{split}
\mathbb{E}\Big\{|\text{GN}_{t}|^2 \Big\} = \sum\nolimits_{m \in \mathcal{M}_t}|a^*_{mt}|^2
\end{split}
\end{align}
Substituting~\eqref{eq_ds1_f}, \eqref{eq_pc1_f}, and~\eqref{eq_nc1_f} into~\eqref{eq:dap_uatf_expanded} gives~\eqref{eq:sinr_dapcpmr}.
Using~\eqref{eq:d_weight}, the corresponding local quantities are
\begin{align*}
[\mathbf d_t]_m
&=\sqrt{p_t^u}\,b_{mtt}^{\DAZSPMRmath},\\
[\mathbf D_t]_{mm}
&=\sum\nolimits_{k\in\mathcal P_{i_t}\setminus\{t\}}p_k^u
\big(b_{mkt}^{\DAZSPMRmath}\big)^2+W_{mt}^{\DAZSPMRmath}.
\end{align*}
Hence,~\eqref{eq:d_weight} gives the \dLSFD\ coefficient in~\eqref{eq:d_dapcpmr}.

\subsection{Gradient Template for DAZS Combining}
\label{app:gradient}

\setcounter{equation}{0}
\renewcommand{\theequation}{D\arabic{equation}}

For either fading model, write the local selection rate as $ R_{mt}= \log_2 \left(1+ \frac{S_{mt}}{I_{mt}}\right)$.
For an allocation variable $\delta_{m\ell}$,
\begin{equation}
\frac{\partial R_{mt}}{\partial\delta_{m\ell}}
=
\frac{
I_{mt}\,\partial_{\ell}S_{mt}
-
S_{mt}\,\partial_{\ell}I_{mt}
}{
\ln(2)\,I_{mt}(I_{mt}+S_{mt})
}.
\label{eq:rate_derivative}
\end{equation}

\subsubsection{Independent Rayleigh fading}
For \DAZSMR\ combining,
\begin{align}
\partial_{\ell}q_{mt}^{\DAZSMRmath}
= &
\begin{cases}
-\Big(1+\sum_{i\neq i_t}\delta_{mi}\Big),
& \ell=i_t,\\[1mm]
-\delta_{mi_t}, & \ell\neq i_t,
\end{cases}
\label{eq:iid_grad_q_mr}
\\
\partial_{\ell}S_{mt}^{\DAZSMRmath}
= &
p_t^u\gamma_{mt}
\partial_{\ell}q_{mt}^{\DAZSMRmath}.
\label{eq:iid_grad_S_mr}
\end{align}
Using the definitions of $C_{mt}$ and $G_{mi}$ in Section~\ref{subsec:dazs_iid_allocation},
\begin{align}
\partial_{\ell}I_{mt}^{\DAZSMRmath}
&=
C_{mt}
\partial_{\ell}q_{mt}^{\DAZSMRmath}
\nonumber\\
&\quad-
\begin{cases}
\displaystyle
\sum_i\delta_{mi}G_{mi}
+\delta_{mi_t}G_{mi_t},
& \ell=i_t,\\[2mm]
\delta_{mi_t}G_{m\ell},
& \ell\neq i_t.
\end{cases}
\label{eq:iid_grad_I_mr}
\end{align}
For \DAZSPMR\ combining,
\begin{align}
\partial_{\ell}S_{mt}^{\DAZSPMRmath}
=
-p_t^u\gamma_{mt},
\quad
\partial_{\ell}I_{mt}^{\DAZSPMRmath}
=
-C_{mt}-G_{m\ell}.
\label{eq:iid_grad_I_pmr}
\end{align}

\subsubsection{Spatially correlated Rayleigh fading}
Let
\begin{equation}
S_{mt}^{\mathcal X}
=
p_t^u q_{mt}^{\mathcal X}\bar{\gamma}_{mt},
\quad
I_{mt}^{\mathcal X}
=
1+N_{mt}^{\mathcal X}.
\label{eq:corr_SI_template}
\end{equation}
For \DAZSMR\ combining,
\begin{align}
\partial_{\ell}q_{mt}^{\DAZSMRmath}
&= 
\begin{cases}
-\Big(
1+\sum_{j\neq t}\delta_{mj}
\Big),
& \ell=t,\\[1mm]
-\delta_{mt},
& \ell\neq t,
\end{cases}
\label{eq:corr_grad_q_mr}
\\
\partial_{\ell}S_{mt}^{\DAZSMRmath}
&= 
p_t^u\bar{\gamma}_{mt}
\partial_{\ell}q_{mt}^{\DAZSMRmath}.
\label{eq:corr_grad_S_mr}
\\
\partial_{\ell}I_{mt}^{\DAZSMRmath}
&{=}
\begin{cases}
{-}p_t^u\gamma_{mtt}^{\rm eff}
{-}\sum\nolimits_{k\neq t}
p_t^u\delta_{mk}\gamma_{mtk}^{\rm eff},
& \ell=t,\\[3mm]
-p^u_{\ell}\delta_{mt}\gamma_{mt\ell}^{\rm eff},
& \ell\neq t.
\end{cases}
\label{eq:corr_grad_I_mr}
\end{align}
For \DAZSPMR\ combining,
\begin{align}
\partial_{\ell}S_{mt}^{\DAZSPMRmath}
=
-p_t^u\bar{\gamma}_{mt},
\quad
\partial_{\ell}I_{mt}^{\DAZSPMRmath}
=
-p^u_{\ell}\gamma_{mt\ell}^{\rm eff}.
\label{eq:corr_grad_I_pmr}
\end{align}
Finally, the gradient of the relaxed objective in~\eqref{eq:penalty} is obtained by summing~\eqref{eq:rate_derivative} over $t\in\mathcal T_m$ and adding the penalty derivatives. For allocation variable $\delta_{m\ell}$, these are
\begin{align}
\partial_{\ell} f_m
={}&
\sum_{t\in\mathcal T_m}
\frac{\partial R_{mt}}{\partial\delta_{m\ell}}
-
2\chi\lambda_1
[\delta_{m\ell}-\delta_{m\ell}^2]_+
(1-2\delta_{m\ell})
\nonumber\\
&-
2\chi\lambda_2
\Big[
\sum\nolimits_i\delta_{mi}-A+1
\Big]_+ ,
\label{eq:full_pga_gradient}
\end{align}
where the summation index in the last term runs over pilots under independent Rayleigh fading and over locally served UE under spatially correlated Rayleigh fading.

\end{document}